\documentclass[aps,prd,twocolumn,floatfix,nofootinbib,superscriptaddress,tightenlines]{revtex4}
\usepackage{soul}
\usepackage{mathtools}
\usepackage[dvipsnames]{xcolor}
\usepackage{amsmath}
\usepackage{dcolumn}
\usepackage{lipsum}
\usepackage{amssymb}
\usepackage{url}
\usepackage{epsfig}
\usepackage{graphicx}
\usepackage{amsmath}
\usepackage{bm}
\usepackage{setspace}
\usepackage{lscape}
\usepackage{amsthm}
\usepackage{bbold}
\usepackage{dcolumn}
\usepackage{epsfig}
\usepackage{graphics}
\usepackage{graphicx}
\usepackage{longtable}

\usepackage{bm}
\usepackage{xspace}
\usepackage{cancel}
\usepackage{float}
\usepackage{multirow}
\definecolor{darkgreen}{rgb}{0,0.5,0}
\definecolor{purple}{rgb}{0.5,0,0.5}
\definecolor{nblue}{rgb}{0.0,0.0,0.50}
\definecolor{scarlet}{rgb}{1.0,0.2,0}
\definecolor{darkmagenta}{rgb}{0.55, 0.0, 0.55}
\definecolor{darkolivegreen}{rgb}{0.33, 0.42, 0.18}
\definecolor{darkcandyapplered}{rgb}{0.64, 0.0, 0.0}

\usepackage[colorlinks=true, pdfstartview=FitV, linkcolor=purple, citecolor= purple, urlcolor=blue]{hyperref}
\usepackage[normalem]{ulem}

\newcommand{\be}{\begin{equation}}
\newcommand{\tu}{\textcolor{red}{u}}

\newcommand{\f}{\textcolor{blue}{f}}
\newcommand{\fu}{\textcolor{blue}{\bar{f_2}}}
\newcommand{\fd}{\textcolor{blue}{f_1}}

\newcommand{\Mav}{\textcolor{blue}{{AV}}}

\newcommand{\td}{\textcolor{darkcandyapplered}{d}}
\newcommand{\tb}{\textcolor{blue}{b}}
\newcommand{\tc}{\textcolor{darkmagenta}{c}}
\newcommand{\ts}{\textcolor{darkgreen}{s}}
\newcommand{\ee}{\end{equation}}
\newcommand{\bea}{\begin{eqnarray}}
\newcommand{\eea}{\end{eqnarray}}
\newcommand{\beas}{\begin{eqnarray*}}
\newcommand{\eeas}{\end{eqnarray*}}
\newcommand{\nn}{\nonumber}

\newcommand{\MeV}{\text{MeV}} 
\newcommand{\GeV}{\text{GeV}} 

\newcommand{\eqn}[1]{Eq.~(\ref{#1})}
\newcommand{\fig}[1]{Fig.~\ref{#1}}
\newcommand{\tab}[1]{Table~\ref{#1}}

\begin{document}
\title{Elastic Form Factors of Axial-Vector Mesons: A Contact Interaction Exploration}

\author{R. J. Hern\'andez-Pinto}
\email{roger@uas.edu.mx}
\affiliation{Facultad de Ciencias F\'isico-Matem\'aticas, Universidad Aut\'onoma de Sinaloa, Ciudad Universitaria, Culiac\'an, Sinaloa 80000,
M\'exico}

\author{L. X. Guti\'errez-Guerrero}
\email{lxgutierrez@secihti.mx}
\affiliation{SECIHTI-Mesoamerican Centre for Theoretical Physics,
Universidad Aut\'onoma de Chiapas, Carretera Zapata Km.~4, Real
del Bosque (Ter\'an), Tuxtla Guti\'errez, Chiapas 29040, M\'exico}

\author{M. A. Bedolla}
\email{marco.bedolla@unach.mx}
\affiliation{Facultad de Ciencias en Física y Matemáticas, Universidad Aut\'onoma de Chiapas, 
Carretera Emiliano Zapata Km.~8, Rancho San Francisco, Ciudad Universitaria Ter\'an, Tuxtla Guti\'errez, Chiapas 29040, M\'exico}

\author{J. P. Uribe-Ramírez}
\email{juanuribe.fcfm@uas.edu.mx}
\affiliation{Facultad de Ciencias F\'isico-Matem\'aticas, Universidad Aut\'onoma de Sinaloa, Ciudad Universitaria, Culiac\'an, Sinaloa 80000,
M\'exico}

\author{A. Bashir}
\email{adnan.bashir@umich.mx \\ adnan.bashir@dci.uhu.es}

\affiliation{Instituto de F\'isica y Matem\'aticas, Universidad
Michoacana de San Nicol\'as de Hidalgo, Edificio C-3, Ciudad
Universitaria, Morelia, Michoac\'an 58040, M\'exico}
\affiliation{Department of Integrated Sciences and Center for Advanced Studies in Physics, Mathematics and Computation, University of Huelva, E-21071 Huelva, Spain}




\begin{abstract}
We employ a symmetry-preserving treatment of the contact interaction within the coupled formalism of Schwinger-Dyson and Bethe-Salpeter equations to calculate the elastic form factors of axial-vector mesons. In this study, we present the computation of the charge radii, magnetic moments, and quadrupole moments of axial-vector mesons, including those composed of light quarks, heavy quarks or a light and a heavy quark.
Our findings indicate that the electric form factor for axial-vector mesons, like that of vector mesons, crosses zero. Furthermore, this crossing occurs at a lower value for axial-vector mesons than for vector mesons.
The results for vector-axial mesons follow a similar hierarchy in charge radii as observed for S, PS, and V mesons, with radii decreasing as the mass of the dressed quarks increases. 
We also include a term associated with the anomalous magnetic moment in the quark-photon vertex. This term has a noticeable impact on both the axial-vector magnetic moment and quadrupole moment, leading to significant percentage changes in their values.
We compare our results with those obtained from other models whenever available.
\end{abstract}


\maketitle

\section{Introduction}
\begin{figure}[htb]
\vspace{-1cm}
       \centerline{
       \includegraphics[scale=0.3,angle=-90]{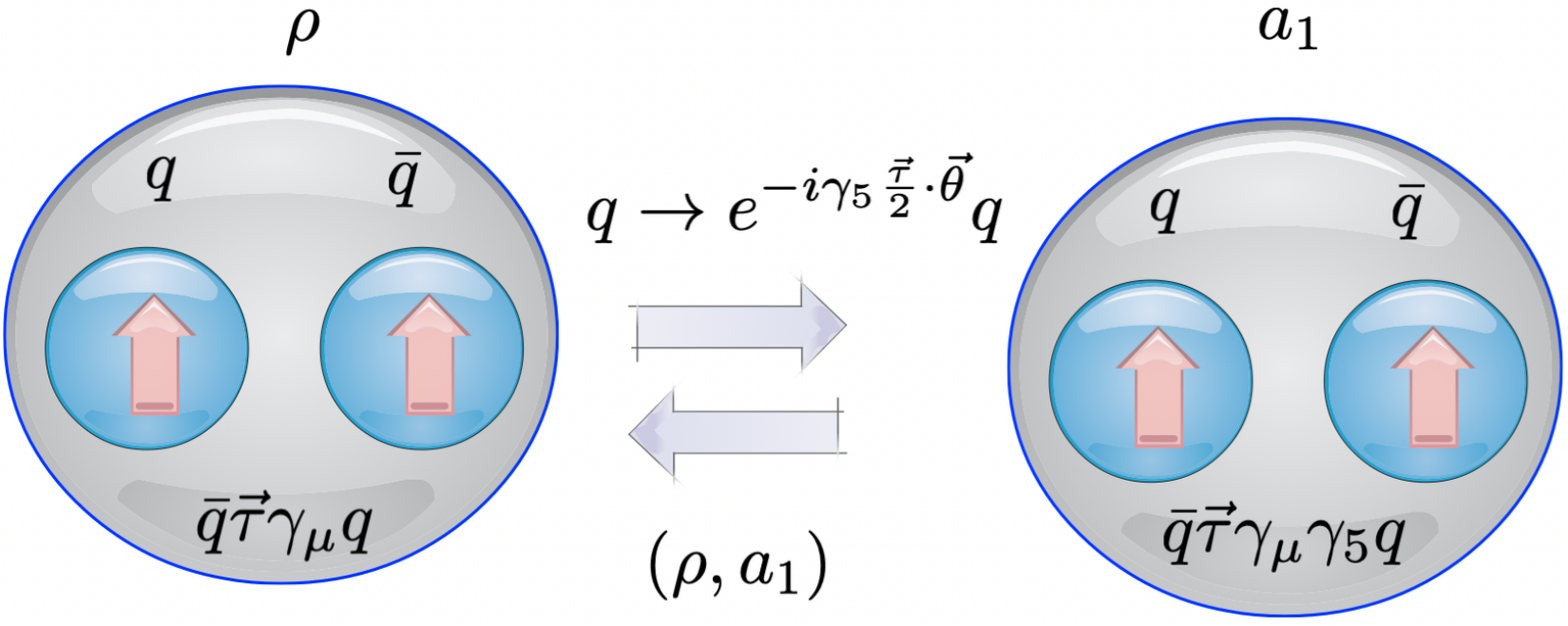}
       }
          \vspace{-1cm}
       \caption{\label{vec-spi} The chiral partners. The $\rho$ and $a_1$ mesons are rotated into each other under chiral transformation. }
\end{figure}

The concept of axial-vector (AV) mesons was introduced early in the development of the quark model \cite{GellMann:1962xb, GellMann:1964nj}. The study of AV mesons is essential for gaining deeper insights into quantum chromodynamics (QCD) and the nature of the strong interactions among quarks and gluons.
AV mesons are characterized by quantum numbers $J^{PC}=1^{++}$ and are the chiral partners of the vector (V) mesons. It implies that they transform into each other under the chiral transformations, forming pairs such as $(\rho, a_1)$ and $(\sigma, \pi)$.
The existence of the lightest AV meson, the $a_1$ meson, was first confirmed by the ACCMOR Collaboration \cite{ACCMOR:1979lrp} through detailed partial-wave analyses of the $\pi^-\pi^-\pi^+$ system. Since this landmark discovery, several additional AV mesons have been observed over the time~\cite{BESIII:2024nhv, DELPHI:2003bnm, TPCTwoGamma:1986cwm, CLEO:1994ait, BaBar:2006eep, LHCb:2015aaf, CMS:2018wcx, BaBar:2014och}, which have greatly enhanced our understanding of hadronic states in the framework of QCD.
While many models have been successful in describing V and pseudoscalar (PS) mesons, scalar (S) mesons and AV mesons remain significant open questions in the field.
One of the key puzzles is understanding the origin of the mass difference between V and AV mesons in the light meson sector. For instance, the mass of the ground state $\rho_{1^-}$ (770 MeV) is notably smaller than that of its chiral partner, the ${a_1}_{1^+}$ (1260 MeV), see~\fig{vec-spi}, with a mass gap of approximately 
500 MeV. This 
puzzle was unveiled by understanding the interplay of dynamical chiral symmetry breaking (DCSB) and the consequent generation of a large
dressed-quark anomalous chromomagnetic 
moment~\cite{Chang:2010hb,Chang:2011ei}. 
It results in the dramatic enhancement of the spin-orbit splitting between ground-state opposite-parity mesons. 

 Form factors of axial vector mesons are relevant for the standard model phenomenology, for example, through contributing to the anomalous magnetic moment $a_\mu$ of the muon~\cite{Aoyama:2020ynm,Cappiello:2019hwh,Roig:2019reh}.
  The AV meson pole diagrams, arising from its transition to two virtual photons, contribute to the hadron light-by-light (HLbL) part of $a_\mu$,  see for example~\cite{Eichmann:2024glq}.
 The AV contributions due to $a_1(1260)$, $f_1(1285)$, and $f_1(1420$) are considered to correspond to about $20\%$ of the total contribution of HLbL \cite{Leutgeb:2019gbz,Aoyama:2020ynm,Eichmann:2024glq,Roig:2019reh,Rudenko:2017bel,Cappiello:2019hwh}.

 Moreover, in the context of strong interactions, investigating the elastic form factors of AV mesons is expected to provide insights into how quarks and gluons interact through QCD to determine their internal structure, including charge distribution, charge radii, magnetic moments, and quadrupole moments. Notably, the electric charge, magnetic moment, and quadrupole moment correspond to the values of their respective form factors at zero momentum transfer.
Despite the intriguing properties of AV mesons, experimental data on their masses, decay constants, and, in particular, their dynamic properties remain limited. We hope that our comprehensive analysis of these states will provide a valuable foundation for future experimental and theoretical investigations.
This study is part of a broader effort to calculate the dynamic properties of all mesons within the framework of a contact interaction (CI) model~\cite{GutierrezGuerrero:2010md,Gutierrez-Guerrero:2019uwa,Gutierrez-Guerrero:2021rsx,Gutierrez-Guerrero:2010waf,Yin:2019bxe,Hernandez-Pinto:2024kwg,Hernandez-Pinto:2023yin,HernandezPinto:2023ric,Wilson:2011aa,Roberts:2011cf,Bedolla:2016yxq}.
In this context, we note that the multipole moments of AV mesons have been calculated using the light-cone sum rules method in~\cite{Aliev:2019lsd}  and within a holographic model in~\cite{Ahmed:2023zkk}.

As previously established, dynamical chiral symmetry breaking induces an anomalous chromomagnetic moment in dressed light quarks. This emergent effect is particularly pronounced at infrared momenta, where its magnitude is comparable to the enhanced magnetic moment of these quarks, albeit with the opposite sign~\cite{Chang:2010hb, Bedolla:2015mpa}.
These intriguing features are crucial for various calculations involving hadronic properties, particularly in the infrared regime. To account for the spin-orbit repulsion, we introduce a phenomenological coupling, 
$g_{SO}=0.25$, as a multiplicative factor in the Bethe-Salpeter kernel, see Refs.~\cite{Roberts:2011cf, Bedolla:2015mpa}. Furthermore, following Refs.~\cite{Wilson:2011aa, Raya:2021pyr,Albino:2025fcp}, we incorporate a term associated with the anomalous magnetic moment in the quark-photon vertex, which can help capture the desirable features of form factors for low momentum transfer.
The implications of introducing quark anomalous magnetic moment for the $\rho$-meson have been explored in~\cite{Xing:2021dwe,Rojas:2024tmn,Xu:2024frc}, but this feature has not yet been investigated in detail for AV mesons. This is one of the objectives of our study.

The article is organized as follows: In Sect.~\ref{BSE}, we outline the essential components required for the analysis within the CI model, including the dressed quark masses obtained by solving the gap equation. We also present the general expressions for the Bethe-Salpeter amplitudes (BSAs) of AV mesons and their numerical results upon solving the Bethe-Salpeter equation. We also include a subsection dedicated to discussing the quark-photon vertex and the introduction of the anomalous magnetic moment term.
In Sect.~\ref{sec:eff}, we examine the general properties of the elastic form factors (EFFs) of AV mesons, focusing specifically on the triangle diagram within the impulse approximation for the 
$M \gamma M$ interaction. This diagram serves as the fundamental building block for calculating meson EFFs in our formalism.
The analytical and numerical results are provided in Subsections~\ref{Analitical} and~\ref{numerical}, respectively. A concise summary and outlook for future research are presented in Sect.~\ref{Summary}. The appendix tabulates  mathematical expressions for the coefficients used in the computation of the EFFs.

\section{The Ingredients} \label{BSE}
Calculation of the meson EFFs presupposes
knowledge of the dynamically generated dressed valence-quark masses, BSAs of the mesons, and the quark-photon interaction vertex at different probing momenta of the incident photon. In this section, we provide a brief but self-contained introduction to the CI, its essential ingredients and characteristics, namely, the gluon propagator, the quark-gluon vertex and the set of parameters employed which, collectively, define the CI. This discussion is followed by the solution of the gap equation to obtain dynamically generated dressed quark masses. We then provide the general expressions of the BSAs for AV mesons. The corresponding  Bethe-Salpeter Equation(BSE) is set up consistently with the gap equation. The numerical solutions are presented in the respective sections dedicated to the analysis of these mesons. The section ends with a detailed discussion of the quark-photon vertex.  

\subsection{The Gap Equation}
\begin{figure}[t!]
   \vspace{-3cm}
   \centering
    \includegraphics[scale=0.4,angle=0]{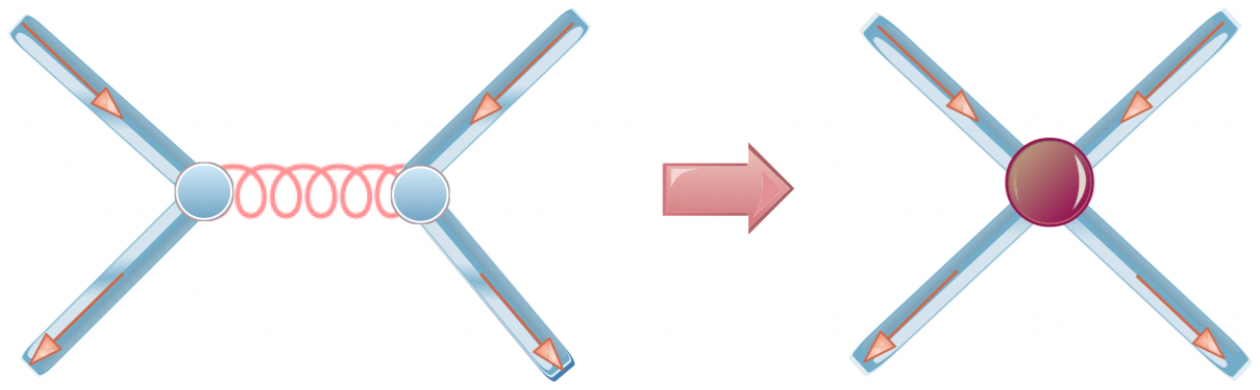}
    \vspace{-4cm}
    \caption{Diagrammatic representation of the CI, employing the simplified model of the gluon propagator in Eq.~(\ref{eqn:contact_interaction}).}
    \label{fig:ci}
\end{figure}

The starting point for our study is the dressed-quark propagator for a quark of flavor $f$,
which is obtained by solving the gap equation,
\begin{eqnarray}
 S(p)^{-1} &=& i\gamma\cdot p + m_{f} + \Sigma(p) \;,\nn \\
\Sigma(p) &=& \frac{4}{3} \int \! \frac{d^4q}{(2\pi)^4} g^2 D_{\mu\nu}(p-q)
\gamma_\mu S(q) \Gamma_\nu(q,p) ,\; \label{gendse}
\end{eqnarray}
where $m_f$ is the Lagrangian level  current-quark mass, $D_{\mu\nu}(p)$ is
the gluon propagator and $\Gamma_\nu(q,p)$ is the quark-gluon vertex. 
It is a well-established fact by now that the Landau gauge gluon propagator saturates in the infrared and a large effective mass scale is generated for the gluon, see for example~\cite{Boucaud:2011ug,Ayala:2012pb,Bashir:2013zha,Binosi:2016nme,Deur:2016tte,Rodriguez-Quintero:2018wma}. It also leads to the saturation of the effective strong coupling at large distances. This modern understanding of infrared QCD
forms the defining ideas of the CI proposed in~\cite{GutierrezGuerrero:2010md}.
We assume that the quarks interact, not through a perturbative massless vector-boson exchange
but via a CI. Thus the gluon propagator no longer runs with a momentum scale but is frozen into a CI in keeping with the infrared properties of QCD, see Fig.~\ref{fig:ci}.
Thus 
\begin{eqnarray}
\label{eqn:contact_interaction}
g^{2}D_{\mu \nu}(k)&=&4\pi\hat{\alpha}_{\mathrm{IR}}\delta_{\mu \nu}  \,,
\end{eqnarray}
\noindent  where $\hat{\alpha}_{\mathrm{IR}}=\alpha_{\mathrm{IR}}/m_g^2$.
The gluon mass scale in QCD, {\em i.e.}, $m_g$ is for dimensional reasons and we take it to be the currently accepted value $m_g=500\,\MeV$~\cite{Aguilar:2017dco,Binosi:2017rwj,Gao:2017uox}. It is clear that in the CI gap equation, the effective coupling which appears is $\hat{\alpha}_{\mathrm{IR}}$ instead of $\alpha_{\mathrm{IR}}$. 
 We choose $\alpha_{\rm IR}/\pi$ to be $0.36$ so that $\hat{\alpha}_{\mathrm{IR}}$ has exactly the same value as in all related previous works~\cite{Gutierrez-Guerrero:2010waf,Gutierrez-Guerrero:2019uwa,Gutierrez-Guerrero:2021rsx,Yin:2019bxe}.
The interaction vertex is bare, i.e.,
$\Gamma_\nu(q,p)=\gamma_\nu$. Note that this model construction has been carried out through our understanding of infrared QCD in the Landau gauge. Gauge-transforming the Green's functions of this simple non-renormaizable model to other gauges is neither instructive nor realistically feasible. However, the fact remains that gauge-covariant truncations of Schwinger-Dyson equations is a desirable pursuit and is carried out in its QCD-akin model building, see for example, Refs.~\cite{Aslam:2015nia,Albino:2018ncl,Albino:2021rvj, Lessa:2022wqc,Ashraf:2025gly}.

This constitutes an algebraically simple but useful and predictive 
rainbow-ladder truncation of the SDE of the quark propagator whose
solution can readily be written as follows:
 \bea\label{DynamicalM}
 S(q,M_f) &\equiv &-i\gamma\cdot q \; \sigma_{V}(q,M_f)+\sigma_{S}(q,M_f)\,,
 \eea
 with
 \bea
 \sigma_{V}(q,M_f)=\frac{1}{q^{2}+M_f^{2}}\, , \sigma_{S}(q,M_f)= M_f
 \, \sigma_{V}(q,M_f) \,, 
 \eea
where $M_f$, for the CI, is the momentum-independent dynamically generated dressed quark mass determined by
\begin{equation}
M_f = m_f + M_f\frac{4\hat{\alpha}_{\rm IR}}{3\pi}
\int_0^\infty \!ds \, s\, \frac{1}{s+M_f^2}\,\,. \label{gap-2}
\end{equation}
Our regularization procedure
 follows Ref.\,\cite{Ebert:1996vx}:
\begin{eqnarray}
\nonumber \frac{1}{s+M_f^2} & = & \int_0^\infty d\tau\,{\rm
e}^{-\tau (s+M_f^2)} \rightarrow \int_{\tau_{\rm UV}^2}^{\tau_{\rm
IR}^2} d\tau\,{\rm e}^{-\tau (s+M_f^2)}
\label{RegC}\\
& = & \frac{{\rm e}^{- (s+M_f^2)\tau_{\rm UV}^2}-e^{-(s+M_f^2)
\tau_{\rm IR}^2}}{s+M_f^2} \,, \label{ExplicitRS}
\end{eqnarray}
where $\tau_{\rm IR,UV}$ are, respectively, infrared and
ultraviolet regulators.  It is apparent from
Eq.\,(\ref{ExplicitRS}) that a finite value of $\tau_{\rm
IR}\equiv 1/\Lambda_{\rm IR}$ implements confinement by ensuring the
absence of quark production thresholds. Since Eq.\,(\ref{gap-2})
does not define a renormalizable theory, $\Lambda_{\rm
UV}\equiv 1/\tau_{\rm UV}$ cannot be removed but instead plays a
dynamical role, setting the scale of all mass dimensioned quantities.
Using Eq.\,\eqref{RegC}, the gap equation becomes
 \bea \label{eq:gapeq}
  M_f = m_f + M_f \frac{4 \hat{\alpha}_{\rm IR}}{3 \pi }
  {\cal C}(M_f^2) \;,
 \eea
 where
 \bea
  \frac{{\cal C}(M^2)}{M^2} = \Gamma(-1,M^2 \tau_{\rm UV}^2) -
  \Gamma(-1,M^2 \tau_{\rm IR}^2) \; 
 \eea
 and $\Gamma(\alpha,x)$ is the incomplete gamma-function.
 
\begin{figure}
    \centering
    \includegraphics[scale=0.74]{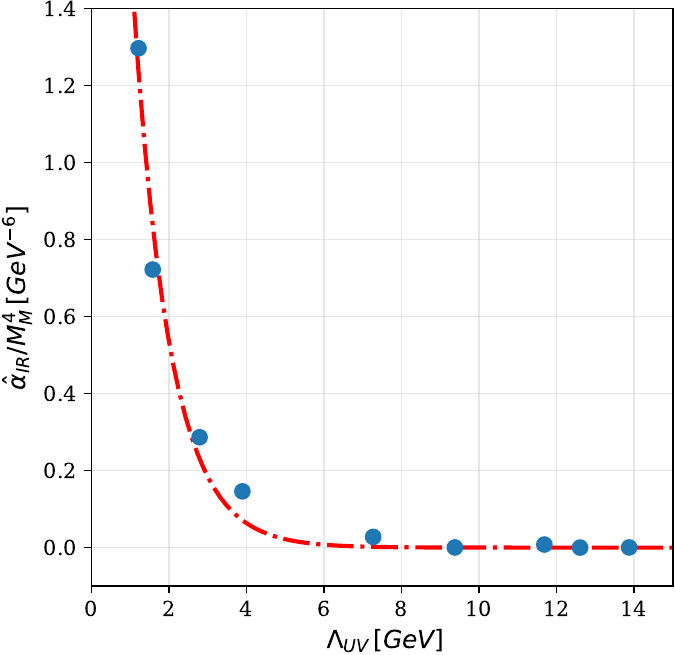}
    \caption{The ratio of the effective coupling to the meson mass is shown as a function of the ultraviolet regulator. The data points correspond to the parameters listed in Table~\ref{parameters}, while the dashed line represents the fit described by \eqn{fitav}.}
    \label{fig:para-av}
\end{figure}

\begin{figure}[t!]
    \centering
    \includegraphics[scale=0.57]{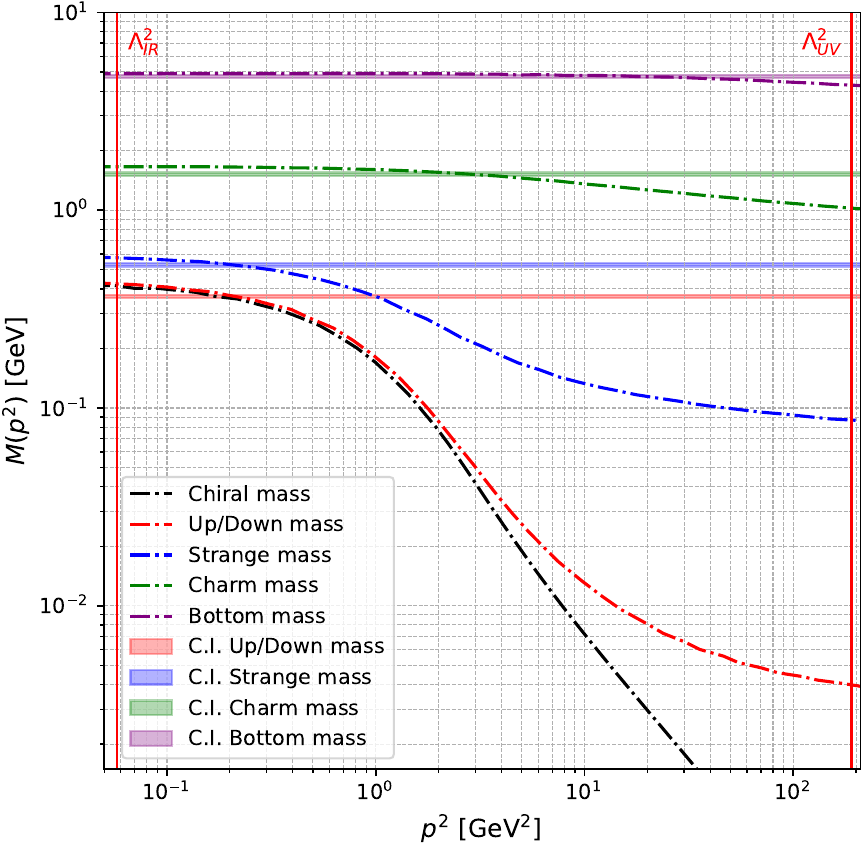}
    \caption{Running dressed quark masses. Dot-dashed lines represents full-QCD lattice results for the mass function~\cite{Bhagwat:2006tu,Serna:2018dwk} while horizontal lines are the constant quark masses generated through the CI approach. $\Lambda_{\rm UV}$ corresponds to the $\tb$-quark value in Tab. \ref{parameters}.}
    \label{fig:running_masses}
\end{figure}

 \begin{table}[ht]
 \caption{ \label{parameters} 
 Ultraviolet regulator and coupling constant for different combinations of quarks in AV mesons.  $\hat{\alpha}_{\mathrm {IR}}=\hat{\alpha}_{\mathrm{IRL}}/Z_H$, where $\hat{\alpha}_{\mathrm {IRL}}=4.57$. $\Lambda_{\rm IR} = 0.24$ GeV is a fixed parameter.} 
\begin{center}
\label{parameters1}
\begin{tabular}{@{\extracolsep{0.0 cm}} || l | c | c | c ||}
\hline \hline
 \, quarks \, &\,  $Z_{H}$ \, &\,  $\Lambda_{\mathrm {UV}}\,[\GeV] $ \,  &\,  $\hat{\alpha}_{\mathrm {IR}}$ \,[\GeV$^{-2}$]
 \\
 \hline
 \rule{0ex}{2.5ex}
$\, \tu,\td, \ts$ & 1 & 1.215 & 4.57   \\ 
\rule{0ex}{2.5ex}
$\, \ts $ & \, 1 \, & 1.580 & 4.57 \\ 
\rule{0ex}{2.5ex}
$\, \tc, \tu $ & \, 0.590 \, & 2.791 & 7.74 \\ 
\rule{0ex}{2.5ex}
$\,  \tc,\ts$ & \, 0.912 \, & 3.895 & 5.01 \\ 
\rule{0ex}{2.5ex}
$\, \tc$     &  1.401 & 7.270 & 3.26 \\
\rule{0ex}{2.5ex}
$\, \tb,\tu$ & 13.686 & 9.378 & 0.33 \\
\rule{0ex}{2.5ex}
 $\, \tb,\ts$ & 0.592 & 11.688 & 7.72 \\
\rule{0ex}{2.5ex}
$\, \tb,\tc$   & 30.972 & 12.610 & 0.15 \\
\rule{0ex}{2.5ex}
$\, \tb$     & 1.426 & 13.876 & \, 3.20 \,  \\
\hline \hline
\end{tabular}
\end{center}
\end{table}
 We report results for all AV mesons  using the parameter values listed in Tables~\ref{parameters},~\ref{table-M}.
The parameters collected in Table~\ref{parameters} indicate that $\hat{\alpha}_{\mathrm {IR}}$ varies with the mass scale $\Lambda_{\mathrm{UV}}$, see Fig.~\ref{fig:para-av}, as described in Ref.~\cite{Bedolla:2015mpa}. It effectively captures the behavior of the effective strong coupling, which should decrease as the energy scale ($\Lambda_{\mathrm{UV}}$) increases. Moreover, as quark masses grow, so does $\Lambda_{\mathrm{UV}}$, representing the highest energy scale of the system under consideration. The red dashed-dotted   curve in Fig.~\ref{fig:para-av} serves as a guide for choosing $Z_{H}$ to ensure a smaller effective $\hat{\alpha}_{\mathrm {IR}}$ at higher $\Lambda_{\mathrm{UV}}$.
This strategy was adopted in several subsequent works~\cite{Bedolla:2016yxq, Raya:2017ggu, Gutierrez-Guerrero:2019uwa, Yin:2019bxe, Yin:2021uom}. 
For the AV mesons, which are the heaviest among  all the mesons studied previously \cite{Pinto:2022tic,HernandezPinto:2023ric,Hernandez-Pinto:2024kwg}, these parameters have been adjusted according to equation ($a$ and $k$ have appropriate dimensions in GeV units):
\bea
\frac{\hat{\alpha}_{\mathrm {IR}}}{M_M^4}=a e^{-k\Lambda_{\mathrm{UV}}},\;\;\; a=4.50 ,\;\;\; k= 1.06\,.
\label{fitav}
\eea
 Nevertheless, the behavior of $\hat{\alpha}_{\mathrm {IR}}$ and $\Lambda_{UV}$ remains consistent with that observed in previous studies: the coupling diminishes (which is in keeping with the fact their electromagnetic sizes reduce) and the ultraviolet cut-off increases as the studied meson is heavier, meaning that the coupling constant and the ultraviolet regulator vary as a function of the meson mass.
\begin{table}[ht]
\caption{\label{table-M}
Current ($m_{f}$) and dressed masses
($M_{f}$) for quarks in GeV, required as an input for the BSE and the EFFs.}
\vspace{0.3cm}
\begin{tabular}{@{\extracolsep{0.0 cm}} || c | c | c | c || }
\hline 
\hline
 $m_{\tu}=0.007$ &$m_{\ts}=0.17$ & $m_{\tc}=1.08$ & $m_{\tb}=3.92$   \\
 \rule{0ex}{2.5ex}
 $ M_{\tu}=0.367$ \, & \, $  M_{\ts}=0.53$\; \, &\,   $  M_{\tc}=1.52$ \, &\,  $  M_{\tb}=4.75$   \\
 \hline
 \hline
\end{tabular}
\end{table}
Table~\ref{table-M} presents the current quark masses $m_f$ used here and the dynamically generated dressed masses $M_f$ of $\tu$, $\ts$, $\tc$ and $\tb$ computed from the gap equation, Eq.~(\ref{eq:gapeq})\footnote{We assume isospin symmetry throughout this work.}. As a result of Eq.~(\ref{eq:gapeq}) the generated dressed-quark masses remains constant through the whole energy range. This behaviour is shown in Fig.~\ref{fig:running_masses} where we compare the theoretical predictions from lattice-QCD~\cite{Bhagwat:2006tu,Serna:2018dwk} and the contact interaction approach. In addition, the dressed quark mass generated in the chiral limit is also presented. As can be seen in the figure, the masses computed with the CI model approach the running masses at $Q^2\to 0$. This is the main reason to use
it to estimate {\it static} properties such as masses and charge radii.

%
 We present the study of all heavy ($Q\bar{Q}$), heavy-light ($Q\bar{q}$) and (review) light ($q\bar{q}$) mesons.
We commence by setting up the BSE for mesons by employing a kernel which is consistent with that of the gap equation to obey axial vector Ward-Takahashi identity and low energy Goldberger-Treiman relations, see Ref.~\cite{Gutierrez-Guerrero:2010waf} for details. 
The solution of the BSE yields BSAs whose general form depends not only on the spin and parity of the meson under consideration  but also on the interaction employed as explained in the next sub-section.

\subsection{Bethe Salpeter Equation}
\label{Bse-av}
\begin{figure}[b!]
   \centering
    \includegraphics[scale=0.5,angle=0]{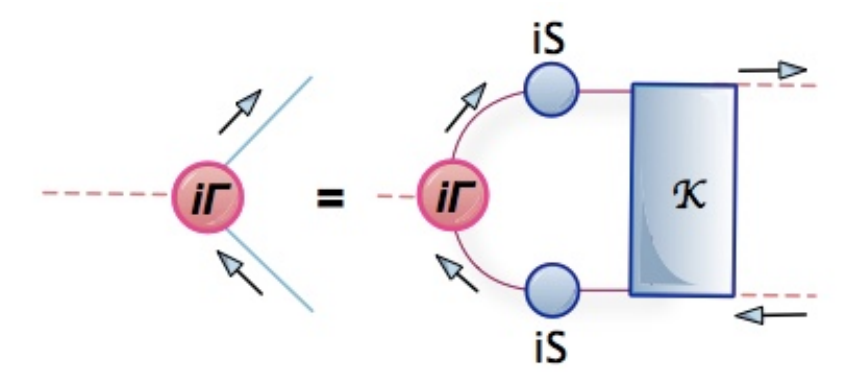}
    \caption{Diagrammatic representation of the BSE. Blue (solid) circles represent dressed quark propagators $S$, red (solid) circle is the meson BSA $\Gamma$ while the blue (solid) rectangle is the dressed-quark-antiquark scattering kernel ${\mathcal {K}}$.}
    \label{fig:BSEfig}
\end{figure}
  The relativistic bound-state problem for hadrons which are characterized by two valence-quarks can be studied using the
 homogeneous BSE whose diagrammatic representation can be seen in  Fig.~\ref{fig:BSEfig}. This equation is mathematically expressed  as follows~\cite{Salpeter:1951sz},
 \begin{equation}
[\Gamma(k;P)]_{tu} = \int \! \frac{d^4q}{(2\pi)^4} [\chi(q;P)]_{sr} {\mathcal K}_{tu}^{rs}(q,k;P)\,,
\label{genbse}
\end{equation}
where $[\Gamma(k;P)]_{tu}$ represents the bound-state's BSA and $\chi(q;P) = S(q+P)\Gamma S(q)$ is the BS wave-function; $r,s,t,u$ represent color, flavor and spinor indices; and ${\mathcal K}$ is the relevant quark-antiquark scattering kernel. This equation possesses solutions on that discrete set of $P^2$-values for which bound-states exist.

A general decomposition of the BSA for the AV mesons ($\fd\fu$) in the CI has the following form
 \bea
\label{AVBSA}
\Gamma^{\Mav}_\mu(P) &=&  \gamma_5 \gamma_\mu^\perp\,E^{\Mav}(P) \,,
\eea
   where $E_{\Mav}(P)$ is the BSA, $P$ is the total meson momentum and  \bea \gamma_{\mu}^{\perp}=\gamma_{\mu}-\frac{ \gamma \cdot P \;
 }{P^{2}} \, P_{\mu}  \,.
 \eea
 Eq.~(\ref{genbse}) has a solution when $P^2=-M_{A}^2$ where $M_{A}$ is the mass of the AV meson. 
We present our results for the masses of the AV mesons studied here in Table \ref{par-AllFF}.  These are composed of different pairs of quarks and antiquarks, corresponding to the light, heavy, and heavy-light sector of the AV mesons.

\begin{table}[h]
\caption{\label{par-AllFF} Calculated values of the BSAs and the masses of the AV mesons by using the parameters in Tables~\ref{parameters} and~\ref{table-M} (compare the parameters with  the ones in Ref.~\cite{Gutierrez-Guerrero:2019uwa}).  \\}  
\begin{tabular}{@{\extracolsep{0.1 cm}} || c | cc | c | c || }
\toprule
 \rule{0ex}{2.5ex}
 &  Mass[GeV]  & $E_{\Mav}$ & $m_{\Mav}^{\rm exp}$[GeV] & error [\%]  \\ 
 \hline
 \rule{0ex}{2.5ex}
 $a_1(\tu\bar\td)$ \,  & 1.373 & 0.32 \, & 1.260 &  8.96 \% \\
 \rule{0ex}{2.5ex}
 $K_1(\tu\bar\ts)$\,   & 1.479 & 0.32 \, & 1.340  &  10.37 \%  \\
\rule{0ex}{2.5ex}
$f_1(\ts\bar{\ts})$ \, & 1.586 & 0.32 \, & 1.430 &  10.90 \% \\
\rule{0ex}{2.5ex}
$D_1(\tc\bar{\tu})$\,  & 2.285 & 0.46 \,& 2.420 &  5.57 \% \\
\rule{0ex}{2.5ex}
$D_{s1}(\tc\bar{\ts})$\, & 2.427 & 0.35 \,& 2.460 &  1.34 \% \\
\rule{0ex}{2.5ex}
$\chi_{c1}(\tc\bar{\tc})$ \, & 3.297 & 0.24 \,& 3.510  & 6.06 \% \\
\rule{0ex}{2.5ex}
$B_1(\tu\bar{\tb})$ \, &  5.612 & 0.12 \,& 5.721 &  1.90 \% \\
\rule{0ex}{2.5ex}
$B_{s1}(\ts\bar{\tb})$\, & 5.583 & 0.50 \,& 5.830 & 4.23\%  \\
\rule{0ex}{2.5ex}
$B_{cb}(\tc\bar{\tb})$\, & 6.521 & 0.06 \,& $\cdots$ &  $\cdots$ \\

\rule{0ex}{2.5ex}
$\chi_{b1}(\tb\bar{\tb})$ \, & 9.608 &  0.16 \, & 9.892 & 2.87 \% \\
\hline
\hline
\end{tabular}
\end{table}

 Measuring electromagnetic form factors involves an incident photon which probes mesons, interacting with the electrically charged quarks making up these two-particles bound states. Therefore, it is natural to look at the structure of the quark-photon vertex within the CI.

\subsection{The Quark-Photon Vertex}
 The electromagnetic probe corresponds to the quark-photon vertex which is denoted by  $\Gamma_{\mu}^{\gamma}(k_+,k_-,M_{\fd})$. It is related to the inverse quark propagator through the following well-known vector Ward-Takahashi
identity:
 \bea
  i P_{\mu} \Gamma_{\mu}^{\gamma}(k_+,k_-,M_{\fd}) =
  S^{-1}(k_+,M_{\fd}) - S^{-1}(k_-,M_{\fd}) \,. \nonumber \\ \label{VWTI}
 \eea
 This identity is essential for a sensible study of a bound-state's EFF. It is determined through the following inhomogeneous BSE,
 \bea
 && \hspace{-1.2cm} \Gamma_{\mu}^{\gamma}(Q,M_{\fd})= \nn \\
 && \gamma_{\mu} - \frac{16 \pi \hat{\alpha}_{\rm IR}}{3} 
  \int  \frac{d^4q}{(2 \pi)^4} \gamma_{\alpha} \chi_{\mu}(q_+,q,M_{\fd})
 \gamma_{\alpha} \, ,\label{eqvertex}
 \eea
where $\chi_{\mu}(q_+,q,M_{\fd})  =  S(q+P,M_{\fd}) \Gamma_{\mu}(Q)S(q,M_{\fd})$.
Owing to the momentum-independent nature of the interaction
kernel, the general form of the solution can simply be written as
  \bea
 && \hspace{-4mm} \Gamma_{\mu}^{\gamma}(Q,M_{\fd})= \gamma_{\mu}^{L}(Q)P_{L}(Q^{2},M_{\fd}) +
 \gamma_{\mu}^{\perp}(Q)P_{T}(Q^{2},M_{\fd}), \nonumber \\
 \eea
 where $\gamma_{\mu}^{L} + \gamma_{\mu}^{\perp} = \gamma_{\mu}$ and
  \bea \gamma_{\mu}^{\perp}(Q)=\gamma_{\mu}-\frac{ \gamma \cdot Q \;
 }{Q^{2}} \, Q_{\mu}  \,.
 \eea
 Inserting this general form into Eq.~(\ref{eqvertex}), one readily
 obtains (note the simplified notation for convenience) the solution for $P_L$ and $P_T$~: 
 \bea
 \hspace{-2mm} P_{L} =1 \,, \quad 
 P_T= \frac{1}{1+K_\gamma(Q^2,M_{\fd})} \,, \label{PTQ2}
 \eea
\vspace{-1cm}
\begin{figure}[htbp]
\centering
\includegraphics[width=9.4cm]{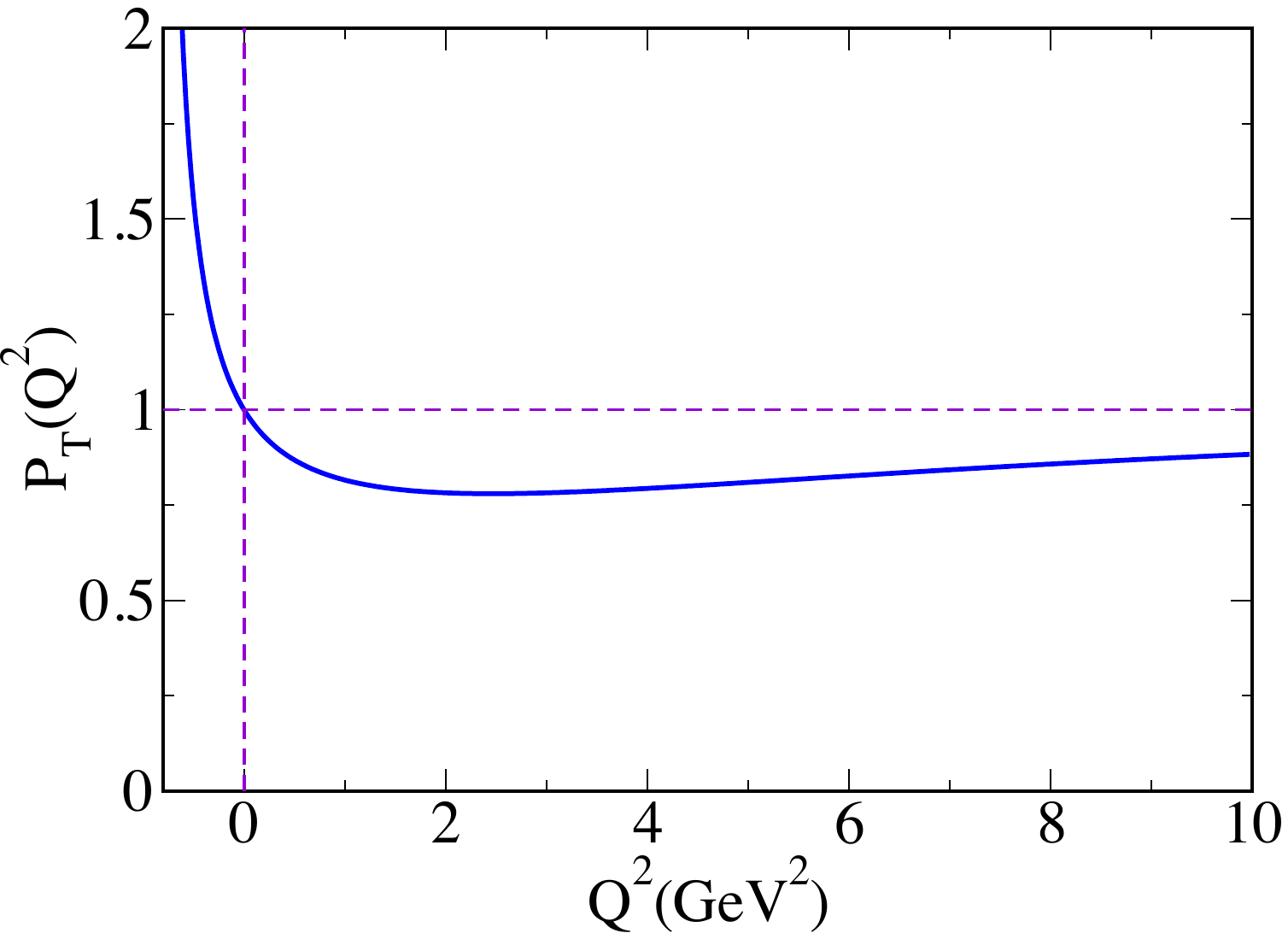}
\caption{Dressing function of the transverse quark-photon vertex, $P_T(Q^2)$, defined in~\protect{Eq.~(\ref{PTQ2}).}}
\label{fig:rhopole}
\end{figure}

\noindent with
\begin{align}
 K_\gamma(Q^2,&M_{\fd}) = \frac{4 \hat{\alpha}_{\rm
IR}}{3\pi} 
\int_0^1d\alpha\, \alpha(1-\alpha) Q^2\,\bar{\mathcal{C}}_1(\omega)
\,,
\end{align}
where
\bea
\bar{\cal C}_1(z) = - \frac{d}{dz}{\cal
C}(z)= \Gamma(0,z\,  \tau_{\rm UV}^2)-\Gamma(0,z \,
\tau_{\rm IR}^2)\,
\eea
and
\bea
 \omega&=&\omega(M_{\fd}^2,\alpha,Q^2) 
 =M_{\fd}^2 + \alpha(1-\alpha) Q^2 \,.
\eea
One can clearly observe from Fig.~\ref{fig:rhopole} that $P_{T}(Q^{2})
\rightarrow 1$ when $Q^2 \rightarrow \infty$, yielding the
perturbative bare vertex $\gamma_{\mu}$ as expected.\\

In addition to the $P_L$ and $P_T$ structures, we now introduce a term associated with the anomalous magnetic moment into the quark-photon vertex through the inhomogeneous Bethe-Salpeter equation \cite{Chang:2010hb}. This approach has been employed in several prior studies of two-quark bound states with $J^P=1^+$ \cite{Wilson:2011aa,Raya:2021pyr,Xing:2021dwe} and the CI study of baryon, see for example~\cite{Raya:2021pyr,Albino:2025fcp}.
We modify the quark-photon vertex as follows
\begin{align}
     \Gamma_\mu^\gamma(Q,M_{\fd}) = \gamma_\mu^\perp P_T(Q^2) + \frac{\xi_{\fd}}{2M_{\fd}}\sigma_{\mu\nu}Q^\nu \exp\left(-\frac{Q^2}{4M_{\fd}^2}\right) .
     \label{VPQ-AM}
 \end{align}
In our case, we use $\xi_{\fd}= 1/2$. 
In~\cite{Raya:2021pyr}, its value was chosen as $1/3$. Note that {\em Beyond Rainbow-Ladder} prediction of
$\xi_{\fd}=0.19$ in ~\cite{Xing:2021dwe} lies within a reasonably broad interval $[0,0.5]$ used in the literature. More recently, it has been varied within the interval $[0,2/3]$ in Ref.~\cite{Albino:2025fcp}. 
The adjustments in the parameters in Table~\ref{parameters}  allow us to keep the value of $\xi_{\fd}=0.5$ constant for all states containing light and  heavy quarks although its flavor-dependence should not be ruled out. It is important to note that there are other instances where it is necessary to go beyond the {\em Rainbow-Ladder approximation}. For example, this is required when evaluating time-like form factors through explicitly incorporating meson degrees of freedom within the Schwinger-Dyson and Bethe-Salpeter equations, see Refs.~\cite{Miramontes:2021xgn,Miramontes:2022uyi,Miramontes:2023ivz}.

 The anomalous magnetic moment has virtually no effect on the elastic
form factors of the S and PS mesons but it does change the
form factors of the AV mesons  noticeably. The work presented in~\cite{Wilson:2011aa} shows that the dressed-quark anomalous magnetic moment increases the magnetic moment by 50\% and enhances the magnitude of the quadrupole moment by 30\% for AV diquarks.
For all these reasons, the vertex in \eqn{VPQ-AM} exhibits the desirable features in our model and is capable of providing us with a reasonably sensible qualitative prediction for the AV meson form factors through a triangle diagram.

Before proceeding, it is important to conduct a reality check. The contact interaction model with enhanced infrared coupling provides only an approximation of the non-perturbative properties of QCD and hadron physics, yet it reproduces meson and baryon masses with sufficient accuracy. Although we extend the model to compute form factors, these results tend to be harder than those predicted by asymptotic (perturbative) QCD. Nevertheless, the model's simplicity enables us to compute these form factors efficiently, providing a valuable benchmark for more precise and accurate calculations when feasible.

\section{\label{sec:eff} Form Factors}

The EFFs provide crucial information on the internal structure of mesons. At low momenta, EFFs allow us to unravel the complexities of non-perturbative QCD, i.e., confinement, dynamical chiral symmetry breaking and the fully dressed quarks. At high energies, we expect to confirm the validity of asymptotic QCD for its realistic models while at intermediate energies, we observe a smooth transition from one facet of strong interactions to the other, all in one single experiment if we are able to chart out a wide range of momentum transfer squared $Q^2$ without breaking up the mesons under study.
While extensive research has been conducted on the electromagnetic form factors of vector mesons in the literature, we focus on the study of axial vector mesons, which remains scarce in the literature to date, in this manuscript. 
All the essential information required for the calculation of the elastic form factors has been gathered in the previous section. Using numerical values of the parameters provided in Tables~\ref{parameters} and~\ref{table-M}, we now proceed to compute the electromagnetic form factors. The next two subsections are dedicated to presenting the detailed analytical expressions and numerical results for axial vector mesons.

\subsection{\label{PS-S-FF} Analytical results}
\label{Analitical}

\begin{figure}[b]
\includegraphics[scale=0.25,angle=0]{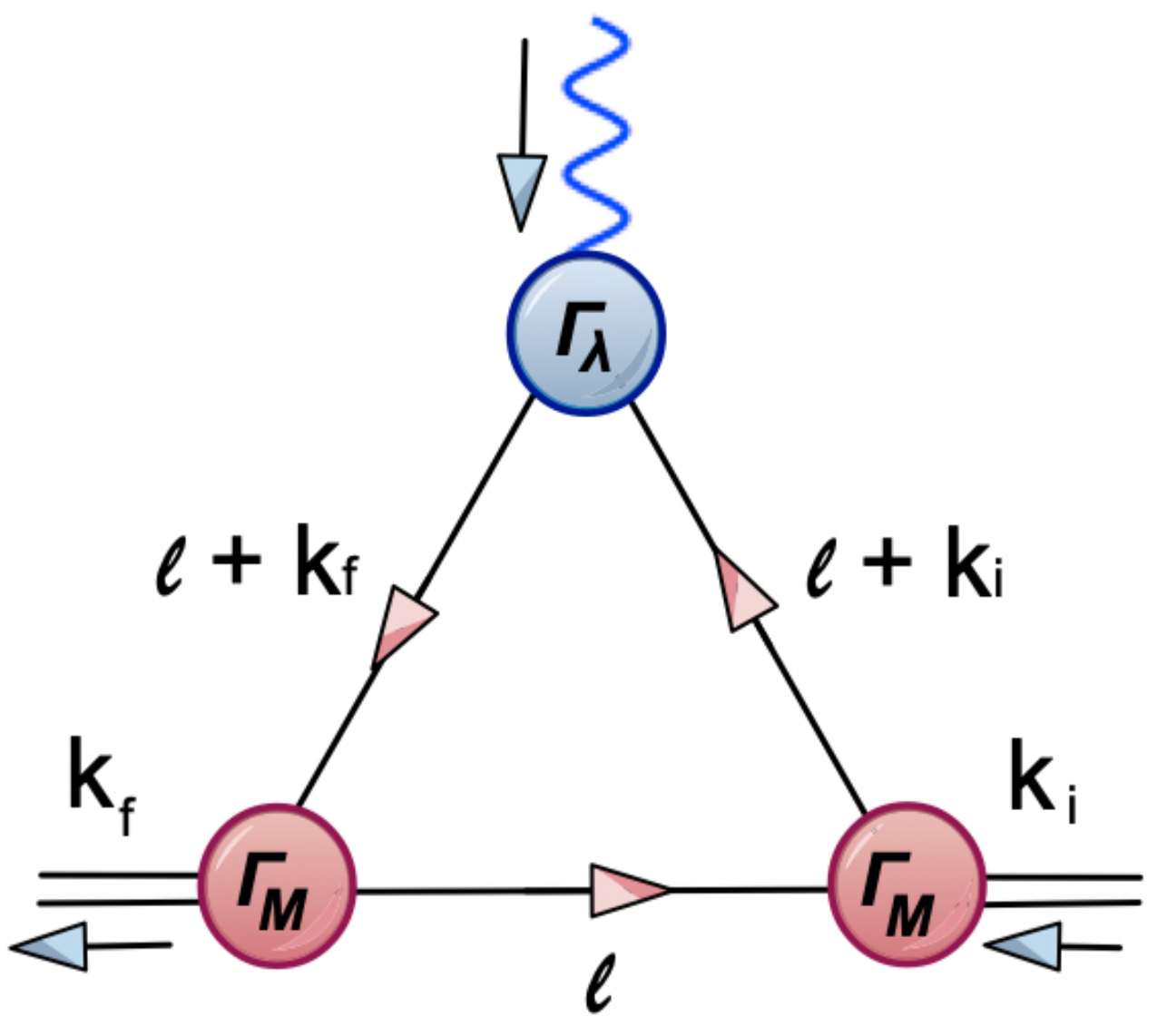}
    \caption{\label{vertex-1} The triangle diagram for the impulse approximation to the $M\gamma M$ vertex.}
\end{figure}

Let us begin with general considerations regarding the electromagnetic interactions of mesons. In the impulse approximation, the $M \gamma M$
vertex, which characterizes the interaction between an AV meson and a photon, is:
\bea
\Lambda_{\mu\nu\lambda}^{\Mav,\fd}&=&N_c\int \frac{d^{4}\ell}{(2\pi)^{4}}
 {\rm Tr}\;\mathcal{G}_{\mu\nu\lambda}^{\Mav,\fd} \,,
 \eea
where
 \begin{align}
 \mathcal{G}^{\Mav,\fd}_{\mu\nu\lambda} &= \nn \, i\bar{\Gamma}^{\Mav}_\mu(k_{f}) \, S(\ell,M_{\fu})  \, i\Gamma^{\Mav}_{\nu}(-k_{i}) \, S(\ell+k_{i},M_{\fd}) \\
  &\times \, i\Gamma^\gamma_{\lambda}(Q,M_{\fd}) 
  \, S(\ell+k_{f},M_{\fd})
 \,. \label{General-FF}
 \end{align}
The notation assumes that it is the quark $\fd$ which interacts with the photon while the antiquark $\fu$ remains a spectator. We  define 
$\Lambda^{\Mav,\fu}$ similarly. Furthermore, we denote the incoming photon momentum by $Q$ while the incoming and outgoing momenta of $M$ by:
$k_{i}=k-Q/2$ and $k_f=k+Q/2$, respectively.
The assignments of momenta are shown in the triangle diagram of Fig.~\ref{vertex-1}.

$\Lambda^{\Mav,\f}$ corresponds to the EFFs of different mesons under study. The contribution from the interaction of the photon with quark $\fd$ can be represented as $F^{\Mav,\fd} (Q^2)$ (stemming from $\Lambda^{\Mav,\fd}$) while the contribution arising from its interaction with quark $\fu$ can be represented as $F^{\Mav,\fu} (Q^2)$ (coming from $\Lambda^{\Mav,\fu}$). The total form factor $F^{\Mav} (Q^2)$ is defined as follows~\cite{Hutauruk:2016sug}:
\begin{equation}\label{eqn:TotalMesonFF}
F^{\Mav} (Q^2) = e_{\fd} F^{\Mav,\fd} (Q^2) + e_{\fu} F^{\Mav,\fu} (Q^2)\,,
\end{equation}
where $e_{\fd}$ and $e_{\fu}$ are the quark and the antiquark electric charges, respectively~\footnote{For neutral mesons composed of same flavored quarks, the total EFF is simply $F^{\Mav} = F^{\Mav,\fd}$.}. For AV, $F^{\Mav,f_1}$ is straightforwardly related to $\Lambda^{\Mav,\fd}$:
\begin{align}\label{Eq:VFF}
     \Lambda_{\lambda \mu \nu}^{\Mav,\fd} = \sum_{j=1}^3 T_{\lambda \mu \nu}^{(j)}(k,Q) \, F_j^{\Mav,\fd}(Q^2) \, ,
 \end{align}with the tensors $T_{\lambda \mu \nu}^{(j)}$ expressed as\,:
\begin{align}
 T_{\lambda \mu \nu}^{(1)}(k,Q) & =  2 k_\lambda\, {\cal
 P}^T_{\mu\alpha}(k_i) \, {\cal P}^T_{\alpha\nu}(k_f)\,, \\
 T_{\lambda \mu \nu}^{(2)}(k,Q) & =  \left[Q_\mu - k_{i\mu} \frac{Q^2}{2 M_{M}^2}\right] {\cal P}^T_{\lambda\nu}(k_f) \nonumber \\
 &- \left[Q_\nu + k^f_\nu \frac{Q^2}{2 M_{M}^2}\right] {\cal P}^T_{\lambda\mu}(k_i)\,, \\ 
 T_{\lambda\mu \nu}^{(3)}(k,Q) & =  \frac{k_\lambda}{M_{M}^2}\, \left[Q_\mu - k_{i\mu} \frac{Q^2}{2 M_{M}^2}\right] \nonumber \\
 &\times \left[Q_\nu + k_{f\nu} \frac{Q^2}{2 M_{M}^2}\right] \,.
 \end{align}
The transverse projector in the above relations is given by the following expression\,:
\begin{eqnarray} {\cal P}_{\alpha
\beta}^T(P) = \delta_{\alpha \beta} - {P_{\alpha}
P_{\beta}}/{P^2} \,.
\end{eqnarray}
An axial vector meson has spin $1$ and positive parity. A parity transformation, relating vector and axial vector mesons,  flips the direction of a particle's motion relative to the direction of its spin.
Despite the fact that axial vector mesons are the least experimentally and phenomenologically studied and constrained, one benefit of the model examined here is its applicability to these puzzling particles.
Similarly to vector mesons~\cite{Hernandez-Pinto:2024kwg}, axial vector mesons contains three $F_i^{AV,f_1}$ components. By applying projector operators, we find these components:
\begin{eqnarray}\label{eqn:vecax}
\nn 
&& \hspace{-1cm} F^{\Mav,\fd}_i(Q^2) = \frac{3}{4\pi^2} (E^{\Mav})^2  \, \int_0^1 d\alpha \, d\beta \, \alpha \, 
 \\ && \times
 \bigg[ \mathcal{A}^{\Mav}_i \, \overline{\mathcal{C}}_1(\omega_2)+(\mathcal{B}^{\Mav}_i -\mathcal{A}^{\Mav}_i \, \omega_2) \, \overline{\mathcal{C}}_2(\omega_2) 
\bigg]\,,
\end{eqnarray}

where the label $i=1,2,3$ corresponds to the three FFs. Moreover, $\mathcal{A}_i^{AV}$ and $\mathcal{B}_i^{AV}$ are tabulated explicitly in the Appendix. 
Therefore, electric, magnetic and quadrupole FFs for AV mesons are defined analogously to V mesons in terms of $F_1, F_2$ and $F_3$ as follows: 
\noindent
\noindent
 \begin{subequations}
\begin{align}
G_E^{\Mav}(Q^2) & =  F_1^{ \Mav}(Q^2)+\frac{2}{3} \eta G_Q^{ \Mav}(Q^2)\,,\\
G_M^{ \Mav}(Q^2) & =  - F_2^{ \Mav}(Q^2)\,, \\
G_Q^{ \Mav}(Q^2) & =  F_1^{ \Mav}(Q^2) + F_2^{ \Mav}(Q^2) + \left[1+\eta\right] F_3^{ \Mav}(Q^2)\,,
\end{align}
\end{subequations}
where $\eta=Q^2/(4 M_{ M}^2)$ with $M_{M}$ being the mass of the AV meson. The values corresponding to the charge, magnetic and quadrupole moments of the AV meson are defined at $Q^2=0$ as,
\begin{subequations}
\begin{eqnarray}
G_E^{ \Mav}(Q^2=0) & = & e_{\Mav}\,, \\
 G_M^{ \Mav}(Q^2=0) & = & \mu_{ \Mav},\;  \\
G_Q^{ \Mav}(Q^2=0) &=& \mathcal{Q}_{ \Mav}\,.
\end{eqnarray}
\end{subequations}
Once the preliminaries are established, we can proceed with the explicit and computationally intensive numerical evaluation of the form factors.

\subsection{\label{PS-S-FF-1}Numerical results}
\label{numerical}
\begin{figure*}[htb]
\begin{tabular}{@{\extracolsep{-2.3 cm}}cc}
\hspace{-2cm}
\includegraphics[scale=0.65]{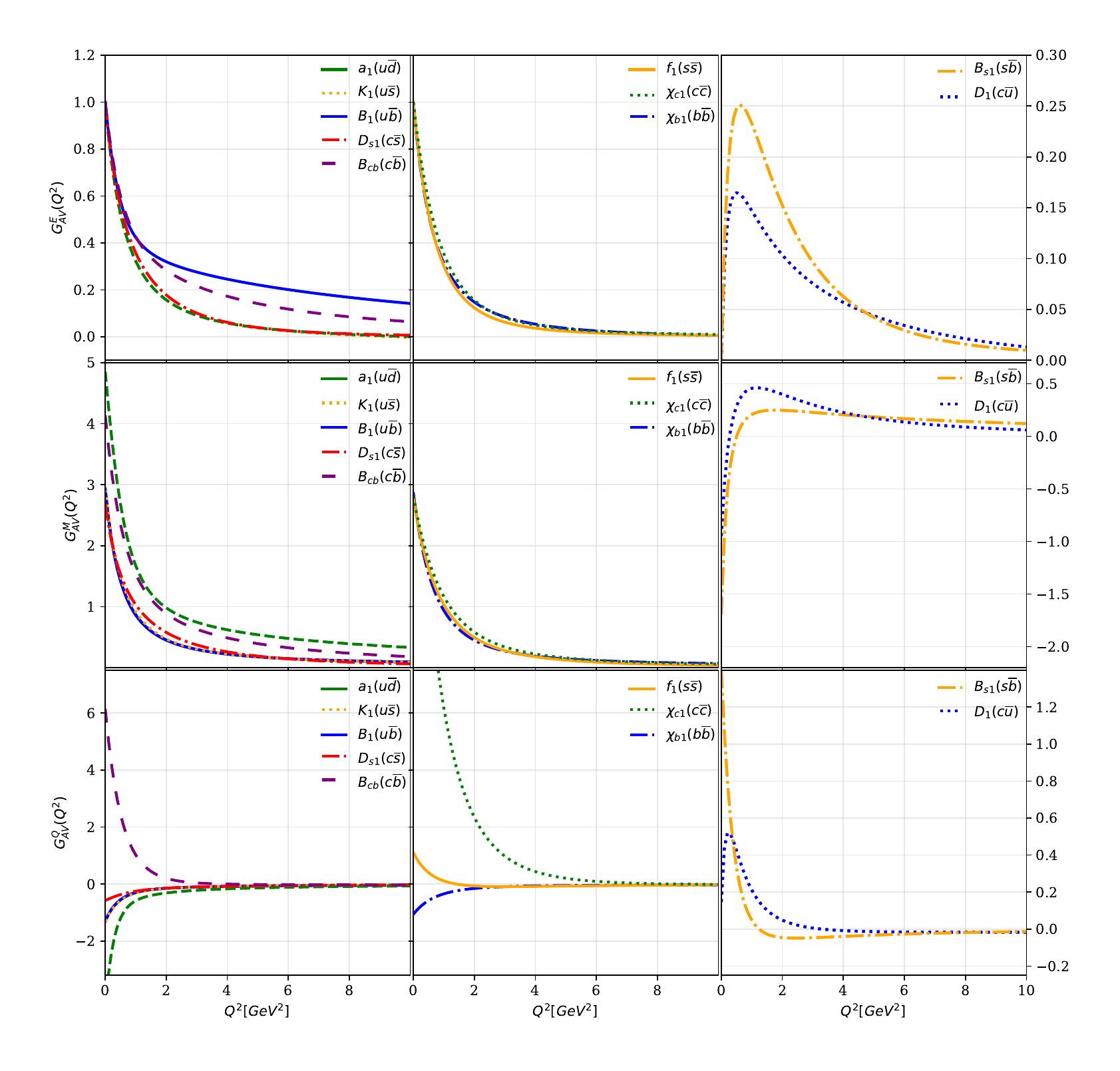}
\end{tabular}
\caption{Electric, magnetic and quadrupole FFs of AV mesons are displayed in the top, middle and bottom rows, respectively. A logarithmic scale is displayed in some plots in order to distinguish the differences. FFs are depicted on the left, center and right columns for charged AV mesons composed of different quarks, neutral AV mesons composed of same flavored quarks and for charged AV mesons, respectively.}\label{plotAE}
\end{figure*}

The predictions of this model for the electric, magnetic and quadrupole FFs for AV mesons are displayed in Fig.~\ref{plotAE}.
We find that the electric form factors of charged mesons decrease more rapidly for lighter mesons in the vicinity of $Q^2=0$, an analogous behavior found for V mesons in Ref. \cite{Hernandez-Pinto:2024kwg}. It occurs similarly for flavor-singlet mesons and, for neutral mesons, the heavier meson increases more rapidly. This observation will be more explicit and physically significant when we compute the charge radii. Moreover, let us recall that for electric form factors, the model predicts that the crossing into negative values occurs for the lightest V mesons at around $Q^2\sim 6$ GeV$^2$.

In Table~\ref{table-Zeros}, we present the locations of the zeros in
$G_{AV}^E(Q^2)$ in terms of $x=Q^2/M_M^2$.
\begin{table}[ht]
\caption{\label{table-Zeros}Comparison of the points where the elastic form factors cross zero for light V and AV mesons.}
\vspace{2mm}
 \renewcommand{\arraystretch}{1.6} %
 \begin{tabular}{@{\extracolsep{0.4cm}}||cc||cc||}
 \toprule
AV & $G^E(Q^2)=0$ & V & $G^E(Q^2)=0$\,  \\
\hline 
  \rule{0ex}{2.5ex}
 $a_1(\tu\bar\td)$& $x=
 4.97$ 
 &$\rho(\tu\bar{\td})$ & $x=6.35$   \\
   \rule{0ex}{2.5ex}
 $K_1(\tu\bar\ts)$ & $x=
 4.65$
 &$K^*(\tu\bar{\ts})$  & $x=6.65$  \\
   \rule{0ex}{2.5ex}
$f_1(\ts\bar{\ts})$ & $x=
4.71$
&$\phi(\ts\bar{\ts})$ & $x=7.59$ \\
\hline \hline
\end{tabular}
\end{table}
It is noteworthy that the point at which $G^E(Q^2)$
crosses zero shifts by approximately $23\%, 32\%, 40\%$ to lower values for axial vector mesons as compared to their vector counterparts. It is due to the fact that we draw the form factors as a function of $x = Q^2/M_M^2$ , where $M_M$ is the meson mass. As axial vector mesons are heavier than their vector counterparts, there is a lower value of $x$ where the corresponding form factors cross zero. Therefor, the zero-crossing is governed by the mass generating mechanism which is largely due to dynamical chiral symmetry breaking for light quarks.

 Experimental data from JLab suggest that the proton’s electric FF might pass through zero at $x\sim 10$~\cite{Jones:1999rz, Gayou:2001qd, Punjabi:2005wq, Puckett:2010ac, Puckett:2011xg}.
 It has been suggested that the likelihood of a zero in the proton elastic electric form factor is 50\% for 
$Q^2 \leq 10.37 \text{GeV}^2$, and this probability rises to 99.99\% for $Q^2\leq 13.06 \text{GeV}^2$~\cite{Cheng:2024cxk}.
 Since the dynamical reasons for the potential appearance of a zero is not expected to be dissimilar in both cases, analyses of $G_V^E(Q^2)$ and $G_{AV}^E(Q^2)$  may provide us with qualitatively guidance on the possible appearance and location of a zero in proton’s EFF.
\begin{figure*}[htbp]
\begin{tabular}{@{\extracolsep{-2.3 cm}}cc}
\hspace{-2cm}
\includegraphics[scale=0.65]{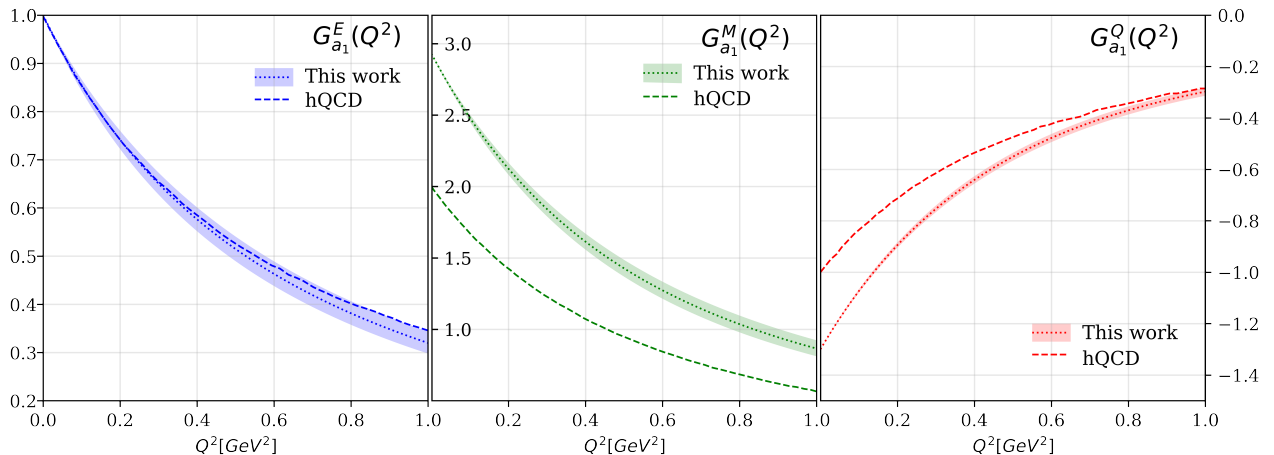}
\vspace{-1cm}
\end{tabular} 
    \caption{Electromagnetic (blue), Magnetic (green), and quadrupole (red) FFs for $a_1$-meson. The central curve in each case is obtained using the $\tau_{\rm UV}$ value from~\tab{parameters1}. The width of the band represents a $5\%$ variation in the charge radius. The meson life is so short that it is challenging to carry out experimental measurements of its EFFs. The graph illustrates a comparison between our results and those obtained using holographic QCD (hQCD) \cite{Ahmed:2023zkk}. } \label{a1}
\end{figure*}
 The magnetic form factors exhibit a distinct variation compared to their respective electric form factors. Near 
$Q^2=0$, the reduction occurs more rapidly for heavier charged and flavor-singlet mesons, whereas for neutral heavier mesons, the increase is more gradual. This behavior is consistent with what has been observed for vector mesons in 
Ref.~\cite{Hernandez-Pinto:2024kwg}.

 Identifying a clear pattern for the quadrupole form factors for AV mesons is challenging, particularly in their behavior as $Q^2 \rightarrow 0$. This contrasts with the V mesons, which consistently tend in the same direction for small exchanged momentum squared~\cite{Hernandez-Pinto:2024kwg}. We examined the sensitivity of the results to slight variations in the choice of parameters, but the overall trend of the form factors remains unchanged.
In Fig.~\ref{a1}, we introduce a 5\% variation in the charge radius derived from the electric form factors through modifying a single parameter, $\tau_{\rm uv}$.
We focus on the electric form factor, as it is the most likely of the three to be determined experimentally, if at all.
We then analyze how this variation propagates into the other two form factors across the entire range of $Q^2$ where calculations have been performed.
See all the plots in
Fig.~\ref{a1} for a comprehensive comparison.
In the future, it will be valuable to move beyond the CI approach by incorporating algebraic models or a full SDE analysis. This would allow for more definitive statements on the detailed behavior of these form factors.

\begin{figure}[ht]
    \centering
    \includegraphics[scale=0.47]{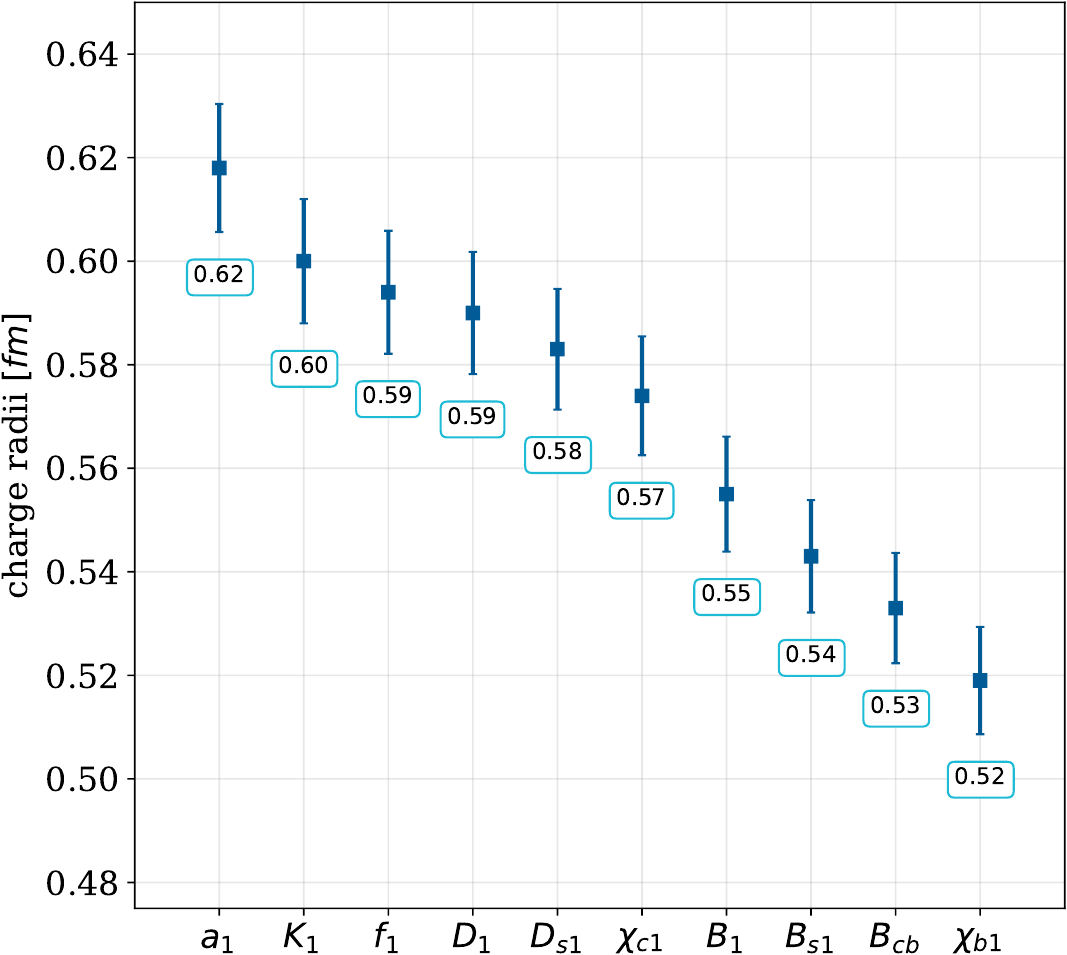}
    \caption{Charge radii of ground state AV mesons in the CI computed using Eq.~(\ref{fradii}). It is clear from this plot that the radii tend to decrease as the  dressed quark masses increase. This significant  behavior is addressed in Sec.~\ref{Summary}.
    }
    \label{jerarquia1}
\end{figure}

\begin{figure}[ht]
\centerline{
\includegraphics[scale=0.25,angle=0]{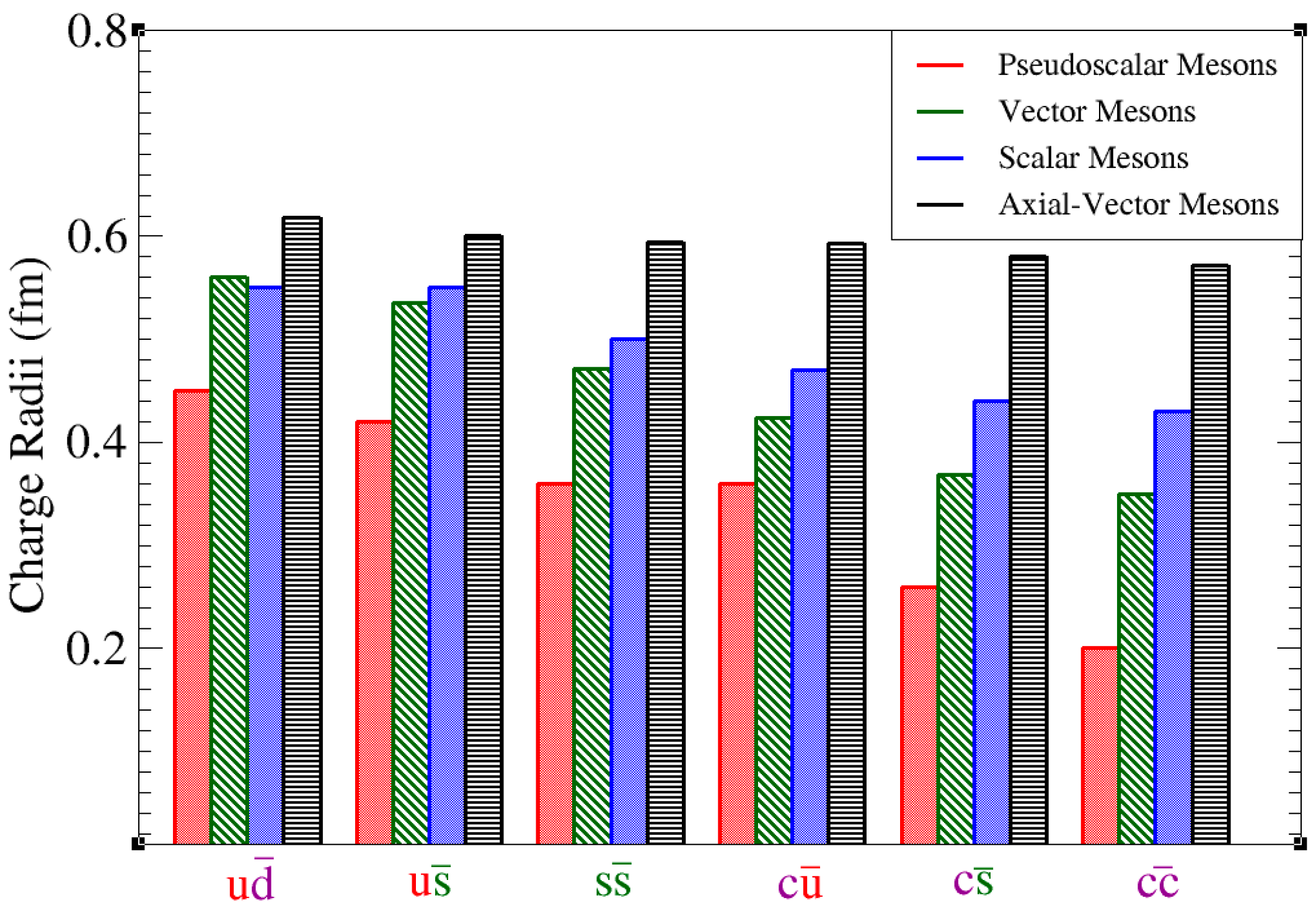}
}
    \caption{\label{jera-charge-1} Charge radii for PS, V, S, and AV mesons composed of $\tu$, 
$\td$, $\ts$, and $\tc$ quarks. The results for the charge radii of V mesons are taken from \cite{Hernandez-Pinto:2024kwg}, and those for S and PS mesons are from \cite{Hernandez-Pinto:2023yin}, using the same formalism employed here for the AV mesons.}
\end{figure}

\begin{figure}[ht]
\centerline{
\includegraphics[scale=0.25,angle=0]{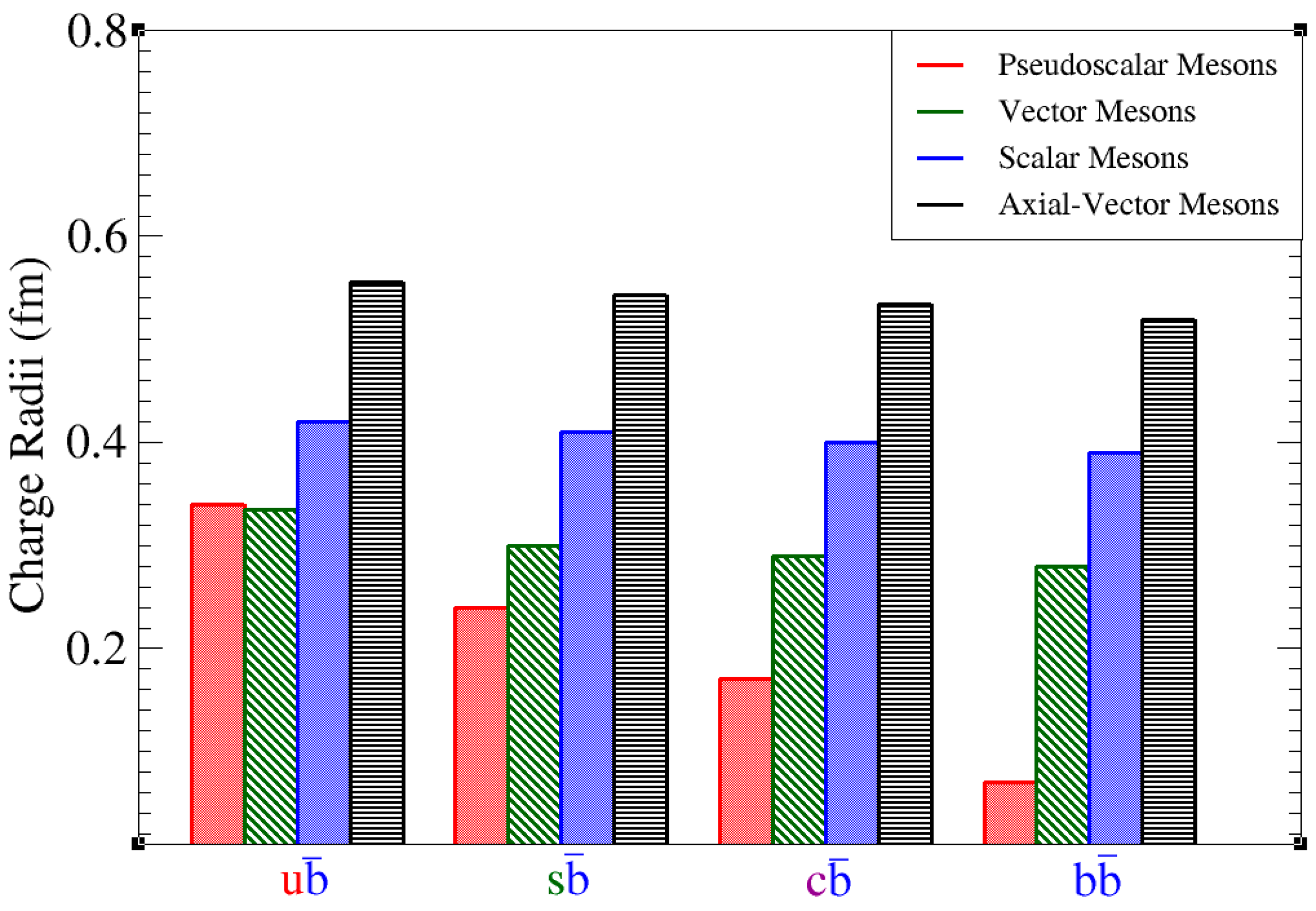}
}
    \caption{\label{jera-charge-2}Charge radii for PS, V, S, and AV mesons composed of $\tu$, $\ts$, $\tc$ and $\tb$ quarks. The results for the charge radii of V mesons are taken from \cite{Hernandez-Pinto:2024kwg}, and those for S and PS mesons are from \cite{Hernandez-Pinto:2023yin}, using the same formalism employed here for the AV mesons.}
\end{figure}

In particular, the results for the elastic form factors of the 
$a_1$-meson are readily comparable to those obtained using holographic QCD (hQCD)~\cite{Ahmed:2023zkk}. The agreement is particularly striking for the electric form factor.
 Following the ideas presented for V mesons, we fit electric, magnetic and quadrupole FFs to the following relations,
\begin{eqnarray}
\nn G_{E}^{\Mav} &=& \frac{e_{\Mav}+a_{\Mav}^E Q^2 + b_{\Mav}^E Q^4}{1+ c_{\Mav}^E Q^2 + d_{\Mav}^E Q^4} \, , \\
\nn G_{M}^{\Mav} &=& \frac{\mu_{\Mav} + a_{\Mav}^M Q^2 +b_{\Mav}^M Q^4}{1+ c_{\Mav}^M Q^2 + d_{\Mav}^M Q^4} \, ,
\end{eqnarray}
\begin{eqnarray}
\nn G_{Q}^{\Mav} &=& \frac{\mathcal{Q}_{\Mav} +a_{\Mav}^Q Q^2 + b_{\Mav}^Q Q^4}{1+ c_{\Mav}^Q Q^2 +d_{\Mav}^Q Q^4} \, .\label{fitsAV}
\end{eqnarray}
We show the values of the coefficients in Table~\ref{tableVEMQA}.  We notice that most $b^i_{AV}$ coefficients, where $i\in \{E,M,Q \}$, are negligibly smaller than $d^i_{AV}$, preserving the observed $1/Q^2$ behavior for the corresponding AV mesons except for those for which $a^i_{AV} \rightarrow 0$ too, leading to an observed $1/Q^4$ fall. The corresponding radii can be calculated as:
\begin{table*}[ht]
 \renewcommand{\arraystretch}{1.6} %
 \caption{\label{tableVEMQA}Parameters for the fits in Eqs.~(\ref{fitsAV}), for the electric magnetic and quadrupole AV FFs.  The form factors for which $b^i_{AV} \rightarrow 0$ exhibit a  $1/Q^2$ behavior for large values of  $Q^2$, with the exception of those for which  $a^i_{AV} \rightarrow 0$ also, depicting a $1/Q^4$ behavior. }
 \vspace{2mm}
\begin{tabular}{@{\extracolsep{0.2 cm}}||ccccccccccccc||} \hline \hline
 & $a_{AV}^E$ &  $b_{AV}^E$ &  $c_{AV}^E$ &  $d_{AV}^E$ & $a_{AV}^M$ &  $b_{AV}^M$ &  $c_{AV}^M$ &  $d_{AV}^M$ & $a_{AV}^Q$ &  $b_{AV}^Q$ &  $c_{AV}^Q$ &  $d_{AV}^Q$   \\ 
\hline
$\tu\bar{\td}$  & 0.196 & $-$0.033 & 1.817& 0.833 &
                  $-$0.128 & 0.102 & 1.200 & 0.999 &
                  $-$0.170 & $-$0.064 &1.999  & 1.992   \\
$\tu\bar{\ts}$  & 0.246 & $-$0.035 & 1.762 & 0.998 
                & 0.056 & 0.072& 1.200 & 0.999 
                & $-$0.256& $-$0.050 & 1.999 & 1.999\\
$\ts\bar{\ts}$  & 0.155 & $-$0.021 & 1.634 & 0.995
                & 0.118 & 0.037 & 1.200 & 0.998 
                & $-$0.054& $-$0.019 & 1.232 & 1.132 \\
$\tc\bar{\tu}$  & 0.192 & $-$0.004 & $-$1.294 & 0.999
                & 1.000 & $-$0.027 & 0.330 & 0.629
                & $-$1.058 & 0.042 & 1.999 & 1.999 \\
$\tc\bar{\ts}$  & $-$0.081 & 0.001 & 1.254 & 0.320 
                & $-$0.023 & 0.001 & 1.200 & 0.298
                & $-$0.061 & 0.001 & 1.242 & 0.309 \\
$\tu\bar{\tb}$  & $-$0.010 & 0.000 & 0.637 & $-$0.001 
                & 0.999 & $-$0.001 & 1.199 & 0.394
                & $-$0.816 & $-$0.002 & 1.999 & 1.999 \\
$\ts\bar{\tb}$  & 0.001 & 0.000 & $-$0.350 & 0.031 
                & 0.787 & $-$0.003 & $-$0.655 & 0.534 
                & $-$0.051 & 0.000 & 0.885 & 0.106 \\
$\tc\bar{\tb}$  & 0.000 & 0.000 & 0.919 & 0.055 
                & 0.249 & 0.000 & 1.199 & 0.245 
                & $-$1.325 & 0.004 & 1.945 & 1.999 \\
$\tc\bar{\tc}$  & $-$0.048 & 0.000 & 1.151 & 0.992
                & 0.025 & 0.000 & 1.020  & 0.678 
                & $-$0.801 & 0.014 & $-$1.099 & 1.797 \\
$\tb\bar{\tb}$  & $-$0.020 & 0.000 & 1.012 & 0.636
                & 0.225 & 0.000 & 0.887 & 0.698
                & $-$1.894 & 0.005 & 1.720 & 1.999 \\ 
\hline\hline
\end{tabular}
\end{table*}
\vspace{-5mm}
\begin{equation}\label{fradii}
 r_{i}^2 =
-6\left.\frac{\mathrm{d}G_{i}(Q^{2})}{\mathrm{d}Q^{2}}\right|_{Q^{2}=0}
\,,
\end{equation}
where $i\in \{E,M,Q\}$. We caution that positive slopes appear in some EFFs; in those cases, the definition for charge radii squared is opted to remove the minus sign in the equation to avoid imaginary numbers. We display our detailed results for the charge, magnetic, and quadrupole radii in Table~\ref{tableradiiAV}, where we also include the corresponding moments, compared with other approaches~\cite{Ahmed:2023zkk,Aliev:2019lsd,Aliev:2009gj,Ozdem:2024qaa,Simonis:2016pnh,Adhikari:2018umb}.
In columns five and seven, we have incorporated the term associated with the anomalous magnetic moment {in our calculations  with the modified quark-photon vertex from \eqn{VPQ-AM}}, whereas in columns six and eight, this term has been omitted. The absolute values of the magnetic and quadrupole moments are generally larger for most AV mesons when anomalous magnetic term is included.
Our results show that the inclusion of this factor in the photon-quark vertex increases the meson magnetic moment by a minimum of 18\% for the $f_1$-meson and by a maximum of 36\% for the $B_{\tc\tb}$-meson.  The enhancement in the quadrupole moment is even more pronounced when this vertex is incorporated.
\begin{table*}[htbp]
\caption{\label{tableradiiAV} Charge radii, magnetic and quadrupole radii (all in fermi)  and moments of AV mesons.  
In the columns corresponding to the magnetic and quadrupole moments, two sets of values are presented: one that incorporates the quark's anomalous magnetic moment (columns five and seven), and another that omits it (columns six and eight, respectively).
The last three columns with the subscript ``o" indicate the values obtained from other models, that is, holographic QCD \cite{Ahmed:2023zkk},
light cone sum rules method \cite{Aliev:2019lsd,Aliev:2009gj,Ozdem:2024qaa},
a bag model \cite{Simonis:2016pnh}, and the basis light front
quantization (BLFQ) approach \cite{Adhikari:2018umb}.}
\vspace{2mm}
 \renewcommand{\arraystretch}{1.6} %
\begin{tabular}{@{\extracolsep{0.4 cm}}||cccc|cc|cc|ccc||}
\hline \hline
 & $r^{E}$ & $r^{M}$ & $r^{Q}$  & \multicolumn{2}{c|}{$\mu$}    &  \multicolumn{2}{c|}{$\mathcal{Q}$} &  $r^{E}_{o}$ & $\mu_{o}$  & $\mathcal{Q}_{o}$ \\ 
\hline
$\tu\bar{\td}$ & 0.618 & 1.088 & 0.772 \, & 2.92 & 2.15  & $-$1.29 &$-0.52$ & 0.62 \cite{Ahmed:2023zkk} & 3.8 \cite{Aliev:2009gj}&$-$0.90 \cite{Aliev:2009gj} \\
$\tu\bar{\ts}$ & 0.600 & 1.038 & 0.753 \,  & 2.91 & 2.19  & $-$1.31 & $-0.45$ & 0.62 \cite{Ahmed:2023zkk}  & 0.28 \cite{Aliev:2009gj} &$-$0.008 \cite{Aliev:2009gj}\\
$\ts\bar{\ts}$ & 0.594 & 0.972 & 0.571 \, & 2.87 & 2.11 & $-$1.06 & $-0.30$ & $\cdots$ & $\cdots$ & $\cdots$ \\
$\tc\bar{\tu}$ & 0.593 & 1.196 & 0.946 \, & $-$0.94 & $-1.35$ \, & 1.39 & $-0.01$ & 0.28 \cite{Ahmed:2023zkk}&0.90 \cite{Aliev:2019lsd}& $-$0.60 \cite{Aliev:2019lsd} \\
$\tc\bar{\ts}$ & 0.581 & 0.946 & 0.330 \, & 2.70 & 2.09  & $-$0.57 & 0.22  &0.32 \cite{Ahmed:2023zkk}&0.87 \cite{Aliev:2019lsd} & $-$0.59 \cite{Aliev:2019lsd} \\
$\tc\bar{\tc}$ & 0.572 & 0.877 & 0.714 \, & 2.79 & 2.05 & 1.10 & 1.84  & 0.27 \cite{Adhikari:2018umb} & $\cdots$ & $\cdots$ \\
$\tu\bar{\tb}$ & 0.555 & 1.135 & 1.769 \, & 4.85 & 6.02 & $-$4.34 & 0.35  & $\cdots$ & 0.60 \cite{Aliev:2019lsd} & $-$0.78 \cite{Aliev:2019lsd} \\
$\ts\bar{\tb}$ & 0.543 & 1.531 & 1.023 \, & $-$1.69 & $-2.09$ \, & 0.14 & $-1.54$\, & $\cdots$ &0.13 \cite{Aliev:2019lsd} & $-$0.10 \cite{Aliev:2019lsd} \\
$\tc\bar{\tb}$ &  0.534 & 1.113 & 1.644 \, & 4.14 & 3.50 & 6.14 & 7.87 & $\cdots$ &$-$0.47 \cite{Ozdem:2024qaa} & $\cdots$ \\
$\tb\bar{\tb}$ & 0.519 & 0.807 & 2.260 \, & 2.81 & 2.07  & 18.53 & 19.28 \, & 0.16 \cite{Adhikari:2018umb} & $\cdots$ & $\cdots$ \\ \hline \hline
\end{tabular}
\end{table*}
The results from the model presented here reveal a clear dependence of the charge radii on the masses of the AV mesons, the larger the mass, the smaller its charge radii, see~\fig{jerarquia1}.
In addition to the hierarchy within mesons of the same type in Table~\ref{tableradiiAV-2}, we also present the charge radii for PS, S, V, and AV mesons. It is readily noticeable that the AV exhibit the largest radii.
The greatest difference between the charge radii of V and AV mesons is found for the 
$\Upsilon$ and $\chi_{\tb_1}$
  mesons, with an 86\%  disparity, while the smallest difference occurs for the $\rho$ and $a_1$
  mesons, with a mere 10\% deviation.
This behavior is easily observed in Figs.~\ref{jera-charge-1}-\ref{jera-charge-2} for the various mesons composed of $\tu,\td,\ts,\tc$ and $\tb$ quarks. 
\begin{table}[ht]
\caption{\label{tableradiiAV-2}
Charge radii (in fermi) of PS, S, and V mesons, obtained in Refs.~\cite{Hernandez-Pinto:2023yin,Hernandez-Pinto:2024kwg}, compared with the ones of AV mesons obtained herein. The last column shows the percentage difference between the radii of the V and AV mesons.} 
\vspace{2mm}
 \renewcommand{\arraystretch}{1.6} %
\begin{tabular}{@{\extracolsep{0.3 cm}}||ccccc|c||}
\hline \hline
 & PS & S & V & AV \,\,  & Diff. (\%) \\
 $\tu\bar{\td}$ & 0.45 & 0.55 & 0.56 & 0.62 \,\, & 10 \\
$\tu\bar{\ts}$ & 0.42 & 0.54 & 0.54 & 0.60 \,\, & 12\\
$\ts\bar{\ts}$ & 0.36 & 0.50 & 0.47 & 0.59 \,\, & 26 \\
$\tc\bar{\tu}$ & 0.36 & 0.47 & 0.42 & 0.59 \,\, & 40  \\
$\tc\bar{\ts}$  & 0.26 & 0.44 & 0.37 &0.58 \,\, & 58 \\
$\tc\bar{\tc}$  & 0.20 & 0.43 & 0.35 & 0.57 \,\,& 63  \\
$\tu\bar{\tb}$ & 0.34 & 0.42 & 0.34 & 0.56 \,\, & 66 \\
$\ts\bar{\tb}$ & 0.24 & 0.41 & 0.30 & 0.54 \,\, & 81 \\
$\tc\bar{\tb}$  & 0.17 & 0.40 & 0.29 & 0.53 \,\, & 84\\
$\tb\bar{\tb}$ & 0.07 & 0.39 & 0.28 & 0.52 \,\, & 86 \\ \hline \hline
\end{tabular}
\end{table}

Finally in Table~\ref{tablemuyq}, we also present a comparison of the magnetic and quadrupole moments for V and AV mesons,  including the ratio between these two moments. It is noticeable that the lighter V mesons are closer to the value of  $\mu/\mathcal{Q}=-2$ expected of {\em structureless} V mesons \cite{Hernandez-Pinto:2024kwg,Brodsky:1992px}. For AV mesons, the $a_1$-meson has a value $\mu/\mathcal{Q}=-2.26$  which differs by 13\% from the value of $-$2. It is straightforward to observe that the value found for the $a_1$-meson is closer to $-$2 than that for $\rho$-meson.
\begin{table}[ht]
\caption{\label{tablemuyq}  Magnetic and quadrupole moments of V and AV mesons. Columns six and seven display the ratio between the magnetic and quadrupole moments.  The only AV mesons that exhibit a sign change in their quadrupole moment compared to their V counterparts are the $\chi_{c1}$, $B_{\tc\tb}$ and $\chi_{b1}$  mesons.
}
\vspace{2mm}
 \renewcommand{\arraystretch}{1.6} %
\begin{tabular}{@{\extracolsep{0.1cm}}||ccc|cc|cc||}
\hline \hline
 &  \multicolumn{2}{c|}{$\mu$}    &  \multicolumn{2}{c|}{$\mathcal{Q}$}  &  \multicolumn{2}{c|}{$\mu$/$\mathcal{Q}$} 
 \\
\hline
& V & AV & V & AV & V & AV 
\\
$\tu\bar{\td}$ & 2.11 & 2.92 & $-$0.85 & $-$1.29 & $-$2.48 & $-$ 2.26 
\\
$\tu\bar{\ts}$ & 2.18 & 2.91 & $-$0.90  & $-$1.31 & $-$2.42 &$-$2.22 
\\
$\ts\bar{\ts}$  & 2.09 & 2.87 & $-$0.83 & $-$1.06 & $-$ 2.51 &$-$2.70 
\\
$\tc\bar{\tu}$ & $-$1.51 & $-$0.94 & 1.05 & 1.39 &$-$1.43& $-$0.67 
\\
$\tc\bar{\ts}$ & 2.10 & 2.70 &$-$0.71 & $-$0.57 & $-$2.95 &$-$4.73 
\\
$\tc\bar{\tc}$ & 2.03&  2.79 &$-$0.70  & 1.10 & $-$2.90 & 2.53 
\\
$\tu\bar{\tb}$ & 3.80 &  4.85 & $-$2.06  & $-$4.34 & $-$1.84 & $-$1.11 
\\
$\ts\bar{\tb}$ & $-$1.82 & $-$1.69 &1.20 & 0.14 &$-$1.51 &$-$12.07 
\\
$\tc\bar{\tb}$ & 2.95 & 4.14 & $-$1.37  & 6.14 & $-$2.15 &  0.67 
\\
$\tb\bar{\tb}$ & 2.01 & 2.81 & $-$0.69  & 18.53 & $-$2.91 & 0.15 
\\ \hline \hline
\end{tabular}
\end{table}

\section{ SUMMARY AND PERSPECTIVE}
\label{Summary}
Our study of the elastic form factors for a set of ten AV mesons $a_1(\tu\bar\td)$,
$K_1(\tu\bar\ts)$,
$f_1(\ts\bar{\ts})$,
$D_1(\tc\bar{\tu})$,
$D_{s1}(\tc\bar{\ts})$,
$\chi_{c1}(\tc\bar{\tc})$,
$B_1(\tu\bar{\tb})$,
$B_{s1}(\ts\bar{\tb})$,
$B_{cb}(\tc\bar{\tb})$,
 and $\chi_{b1}(\tb\bar{\tb})$, composed of both light and heavy quarks, using the contact interaction model, yields the following key findings:
 \begin{itemize}
     \item We calculate the electric, magnetic, and quadrupolar form factors for the AV mesons. The results are presented in \fig{plotAE}. The three form factors exhibit a $1/Q^2$ behavior at large values of $Q^2$, except for 
     a few with $1/Q^4$ fall-off.
    \item In the case of the lightest AV meson,  allowing for a 5\% variation in the charge radius,  we compare it with the results obtained using hQCD~\cite{Ahmed:2023zkk}, finding good agreement between the two.
     \item We present the fits for all the elastic form factors of the AV mesons in \eqn{fitsAV}, using the parameters provided in Table \ref{tableVEMQA}.
    \item We compute the charge, magnetic, and quadrupole radii, along with the corresponding moments, which are presented in Table \ref{tableradiiAV}. These results are compared with those obtained using holographic QCD \cite{Ahmed:2023zkk}, light-cone sum rules \cite{Aliev:2019lsd, Aliev:2009gj, Ozdem:2024qaa}, a bag model \cite{Simonis:2016pnh}, and the basis light front quantization (BLFQ) approach~\cite{Adhikari:2018umb}.
    \item We have introduced a term that incorporates the anomalous magnetic moment in the quark-photon vertex of~\eqn{VPQ-AM}, leading to an increase of 18\% to 30\% in the magnetic moments, which is in good agreement with the results obtained in~\cite{Wilson:2011aa} for AV diquarks. This is presented in Table \ref{tableradiiAV}, where we compare the results with and without the inclusion of this term.
    \item Our results for the charge radii show good agreement with the hierarchy observed for PS, V, and S mesons, where
\vspace*{-3mm}
    \begin{eqnarray}
&& r_{\tu\bar{\td}} > r_{\tu\bar{\ts}} > r_{\tc\bar{\tu}} > r_{\tu\bar{\tb}} \,, \nn \\
&& r_{\tu\bar{\ts}} > r_{\ts\bar{\ts}} > r_{\tc\bar{\ts}} > r_{\ts\bar{\tb}} \,, \nn \\
&& r_{\tc\bar{\tu}} > r_{\tc\bar{\ts}} > r_{\tc\bar{\tc}} > r_{\tc\bar{\tb}} \,, \nn \\
&& r_{\tu\bar{\tu}} > r_{\ts\bar{\ts}} > r_{\tc\bar{\tc}} > r_{\tb\bar{\tb}} \,.\nn 
\end{eqnarray}
\vspace*{-2mm}
This behavior is also graphically shown in \fig{jerarquia1}.
    \item  Our analysis also demonstrates that within the CI, the charge radii of AV mesons are the largest compared to those of other types of mesons, as shown in~\fig{jera-charge-1} -~\fig{jera-charge-2} and Table~\ref{tableradiiAV-2}.
    \begin{eqnarray}
        && r^E_{AV} > r^E_{S} >  r^E_{V} >  r^E_{PS} \,, \nn
    \end{eqnarray}
    Masses of the AV mesons are also the largest~\cite{Gutierrez-Guerrero:2021rsx}.
    \item We compare the magnetic and quadrupole moments of V and AV mesons in Table \ref{tablemuyq}, where we also present the ratio between these two values. It is well known that for structureless spin-one mesons, this ratio should be equal to $-$2 \cite{Brodsky:1992px}. For the lightest AV meson, this value is $-$2.26, differing from $-$2 by 13\%, which is closer than the result obtained for the $\rho$-meson.
    \item The work presented here also highlights the location at which the electric form factor crosses zero, comparing them with those of the V mesons in Table~\ref{table-Zeros}. We emphasize that the zero-crossing point of the $G^E$ decreases for AV mesons compared to their V counterparts.
 \end{itemize}
 
This study on the elastic form factors of axial-vector mesons marks the culmination of an extensive investigation into the elastic form factors of pseudoscalar and scalar mesons~\cite{HernandezPinto:2023ric,Hernandez-Pinto:2023yin}, as well as vector (V) mesons~\cite{Hernandez-Pinto:2024kwg}, all composed of both heavy and light quarks. Additionally, it paves the way for future research into these form factors, either through more sophisticated algebraic models or a QCD-inspired analysis based on the Schwinger-Dyson and Bethe-Salpeter equations. 

We acknowledge the need for experimental insights to explore the key characteristics of axial-vector meson form factors. Currently, valuable data from electron-positron colliders provide crucial information on processes such as: 
$e^+ e^-   \rightarrow  e^+ e^- f_1 $,
$e^+ e^-   \rightarrow  \pi^+ \pi^- f_1 $.
These results impose stringent constraints on the high-energy behavior of these reactions, enabling the determination of the couplings associated with the corresponding transition form factors. However, the direct measurement of electromagnetic form factors is likely to remain an elusive challenge for the foreseeable future. 

Our calculation represents an initial step toward computing the electromagnetic and transition form factors of baryons within the contact interaction model. The results presented here for axial-vector mesons are directly connected to those of vector diquarks. These diquarks play a crucial role in the calculation of baryon form factors within the quark-diquark framework, a key area of investigation for major laboratories. Therefore, baryon studies in experiments can provide us with indirect probe to investigate the electromagnetic form factors of axial vector mesons.

\vspace*{-4mm}
\begin{acknowledgements}
\vspace*{-2mm}
L.~X.~Guti\'errez-Guerrero acknowledges the {\em Secretaría de Ciencia, Humanidades, Tecnología e Inovación} (SECIHTI) for the support provided to her through the {\em Investigadores e Investigadoras por México} SECIHTI program and Project CBF2023-2024-268, Hadronic Physics at JLab: Deciphering the Internal Structure of Mesons and Baryons, from the 2023-2024 frontier science call. The work of R.~J.~Hern\'andez-Pinto is partly supported by SECIHTI (Mexico) through {\em Sistema Nacional de Investigadores}.
 A.~Bashir wishes to acknowledge the {\em Coordinaci\'on de la Investigaci\'on Cient\'ifica} of the{\em Universidad Michoacana de San Nicol\'as de Hidalgo}, Morelia, Mexico,  grant no. 4.10, SECIHTI (Mexico), project CBF2023-2024-3544
as well as the Beatriz-Galindo support during his scientific stay at the University of Huelva, Huelva, Spain. We thank K.~Raya for helpful comments on the draft version of this article. 
\end{acknowledgements}
\vspace*{5mm}
\section*{Appendix: Form Factor Formulae}
\label{App:EM}

In this appendix, we provide the analytic expressions for all coefficients required to determine the axial-vector (AV) electromagnetic form factors (EFFs) within the contact interaction (CI) model.
 In order to add the anomalous magnetic moment piece to the quark-photon vertex in an adequate manner, which probes the electromagnetic form factors of the axial vector mesons through the interaction $M_{AV} \, \gamma \, M_{AV}$, we define 
\begin{align}
\zeta_{\fd}=\frac{\xi_{\fd}}{2M_{\fd}} \exp\left(-\frac{Q^2}{4M_{\fd}} \right) \;,   
\end{align}
 and find the expressions,
 \begin{align}
   \mathcal{A}^{\Mav}_i= \mathcal{A}^{P}_i \, P_T(Q^2) + \mathcal{A}_i \, \zeta_{\fd}  \, , \\
   \mathcal{B}^{\Mav}_i= \mathcal{B}^{P}_i \, P_T(Q^2) + \mathcal{B}_i \, \zeta_{\fd} \, ,
 \end{align}
where $\mathcal{A}_1 =\nn\mathcal{A}_3 = 0$, $\mathcal{A}_2 = -4M_{\fd}$, 
\begin{align}
\nn\mathcal{A}^{P}_1 &= 2-\alpha \,\,\,  , \mathcal{A}^{P}_2 = -2 - 2\alpha(1 - \beta) \, ,\\
\nn \mathcal{A}^{P}_3 &= \frac{4\alpha(1-2\beta)M_M^2}{Q^2+4 M^2_{M}} \, ,  
\end{align}
 with
\begin{align}
\nn\mathcal{B}_1^{P} &=  2 (\alpha M_{\fd}^2 - 2 ( 1 -\alpha )M_{\fd} M_{\fu} + (1 - \alpha)^2 \alpha M_M^2 \\
   \nn &-  (2 - \alpha)  (1 - \beta) \alpha^2 \beta Q^2 ) , \\
\nn\mathcal{B}_2^{P} &= 4 (M_{\fd} M_{\fu} -  \alpha  (1 - \beta) M_{\fd}^2 + (1 - \beta)^2 \alpha^3 \beta  Q^2 \\
   \nn &- (1 - \alpha)  (\alpha + (1-\alpha)
\beta) \alpha M_M^2) \,,\\
\nn\mathcal{B}_3^{P}&= \frac{8 \alpha M_M^2 }{4 M_M^2 + Q^2}\left[   (1 - 2 \beta) M_{\fd}^2 \right. \\
\nn &+   (1 - \beta)  (2 - \alpha (3 -  2 \beta)) \alpha\beta Q^2 \\
\nn &\left. -  (1 - \alpha) (1 - 2 \beta - \alpha (1 + 6 \beta - 8 \beta^2)) M_M^2 \right] \,, 
\end{align}
and
\begin{align}
\nn\mathcal{B}_1 &=2  ( \alpha M_{\fd} -  (1 - \alpha) M_{\fu} )  Q^2 \,, \\
\nn\mathcal{B}_2 &= 4 (M_{\fd}^2 M_{\fu} + (1 - \alpha)^2 M_{\fu} M_M^2 - 2  (1 - \alpha) \alpha M_{\fd} M_M^2 \\ 
   \nn & - \alpha
    (M_{\fu} -  \alpha M_{\fd} + 
       \alpha (-1 + \beta)M_{\fu} ) (1 - \beta) Q^2  ) \,, \\
\nn\mathcal{B}_3 &= \frac{8 M_M^2( \alpha M_{\fd} -  (1 - \alpha)M_{\fu} ) }{4 M_M^2 + Q^2} \\
 &\times (4  (1 - \alpha) M_M^2 + 
    (1 - 2 \alpha (1 - \beta))Q^2) \,. 
\end{align}

\newpage

\bibliography{ccc-a}

\end{document}